\documentclass[twocolumn,byrevtex,prd,reprint,nofootinbib,superscriptaddress]{revtex4-1}   
\usepackage{graphicx}
\usepackage{amsmath}
\usepackage{amsfonts}
\usepackage{amssymb}
\usepackage{bm}
\usepackage{color}
\usepackage[x11names,dvipsnames]{xcolor}
\usepackage[colorlinks=true,allcolors=BlueViolet]{hyperref}

\newcommand\bsub{\begin{subequations}}
\newcommand\esub{\end{subequations}}

\usepackage{xspace}
\usepackage{slashed}
\usepackage{gensymb}

\newcommand{\cf}{{\it cf.}\xspace}

\newcommand{\diff}{\textrm{d}}
\newcommand{\ket}[1]{\left\lvert  #1 \right\rangle}
\newcommand{\bra}[1]{\left\langle #1 \right\rvert }

\newcommand{\ie}{{\it i.e.}\xspace}

\DeclareMathOperator{\im}{Im}
\DeclareMathOperator{\re}{Re}

\usepackage{soul,color}
\definecolor{chromeyellow}{rgb}{1.0, 0.65, 0.0}
\definecolor{applegreen}{rgb}{0.55, 0.71, 0.0}
\definecolor{asparagus}{rgb}{0.53, 0.66, 0.42}

\begin{document}

\title{\boldmath Moments of angular distribution and beam asymmetries in $\eta\pi^0$ photoproduction at GlueX}

\author{V.~Mathieu}
\email{vmathieu@ucm.es}
\affiliation{Theory Center, Thomas  Jefferson  National  Accelerator  Facility, Newport  News,  VA  23606,  USA}
\affiliation{Departamento de F\'isica Te\'orica, Universidad Complutense de Madrid, 28040 Madrid, Spain}

\author{M.~Albaladejo}
\email{albalade@jlab.org}
\affiliation{Theory Center, Thomas  Jefferson  National  Accelerator  Facility, Newport  News,  VA  23606,  USA}

\author{C.~Fern\'andez-Ram\'irez}
\affiliation{Instituto de Ciencias Nucleares, Universidad Nacional Aut\'onoma de M\'exico, Ciudad de M\'exico 04510, Mexico}

\author{A.~W.~Jackura}
\affiliation{Center for  Exploration  of  Energy  and  Matter, Indiana  University,Bloomington,  IN  47403,  USA}
\affiliation{Physics  Department, Indiana  University,Bloomington,  IN  47405,  USA}

\author{M.~Mikhasenko}
\affiliation{CERN, 1211 Geneva 23, Switzerland}

\author{A.~Pilloni}
\affiliation{European Centre for Theoretical Studies in Nuclear Physics and related Areas (ECT$^*$) and Fondazione Bruno Kessler, Villazzano (Trento), I-38123, Italy}
\affiliation{INFN Sezione di Genova, Genova, I-16146, Italy}

\author{A.~P.~Szczepaniak}
\affiliation{Theory Center, Thomas  Jefferson  National  Accelerator  Facility, Newport  News,  VA  23606,  USA}
\affiliation{Center for  Exploration  of  Energy  and  Matter, Indiana  University,Bloomington,  IN  47403,  USA}
\affiliation{Physics  Department, Indiana  University,Bloomington,  IN  47405,  USA}

\collaboration{Joint Physics Analysis Center}

\preprint{JLAB-THY-19-2958}

\date{Received: date / Accepted: date}

\begin{abstract}
In the search for exotic mesons, the GlueX collaboration will soon extract moments of the $\eta\pi^0$ angular distribution.
In the perspective of these results, we generalize the formalism of moment extraction to the case in which the two mesons are produced with a linearly polarized beam, and build a model for the reaction $\vec \gamma p \to \eta \pi^0 p$. 
The model includes resonant \\\mbox{$S$-,~$P$-,~$D$-waves} in $\eta\pi^0$, produced by natural exchanges. Moments of the $\eta\pi^0$ angular distribution are computed with and without the $P$-wave, to illustrate the sensitivity to exotic resonances. Although little sensitivity to the $P$-wave is found in moments of even angular momentum, moments of odd angular momentum are proportional to the interference between the $P$-wave and the dominant $S$- and $D$-waves. We also generalize the definition of the beam asymmetry for two mesons photoproduction and show that, when the meson momenta are perpendicular to the reaction plane, the beam asymmetry enhances the sensitivity to the exotic $P$-wave.
\end{abstract}

\maketitle

\section{Introduction}
The recent 12 GeV upgrade of CEBAF at the 
Jefferson Lab (JLab) opens a new area in meson spectroscopy studies especially in addressing the role of gluons in forming exotic hybrid mesons~\cite{Dudek:2012vr}. The golden channel for discovery of the exotic hybrid(s) is through its  decay to  $\eta^{(')}\pi$ final states. In these final states the odd waves have exotic quantum numbers and the lowest of them, the $P$-wave is expected to resonate due to the exotic $\pi_1(1400/1600)$ state. Properties of this resonance were recently determined using data collected by the COMPASS experiment~\cite{Rodas:2018owy}. According to theoretical predictions~\cite{Szczepaniak:2001qz} \cite{Anikin:2004ja} \cite{Anikin:2004vc}, the production of the exotic meson $\pi_1(1400/1600)$ with photons could lead to sizable cross sections measurable at JLab. 
       
In the present paper, we focus on the reaction $\vec\gamma p \to \eta\pi^0 p$, which is currently under study by the GlueX collaboration. The GlueX  experiment~\cite{gluex} uses linearly polarized photons with energy \mbox{$E_\gamma \sim 9$~GeV}. Observables directly related to the spin of the  resonance in the di-meson spectrum are  moments of the  angular distribution. For example, recent analysis  of moments in $\pi^+\pi^-$ \cite{Battaglieri:2009aa} \cite{Battaglieri:2008ps} \cite{Bibrzycki:2018pgu} and $K^+ K^-$  photoproduction \cite{Lombardo:2018gog} were used to constrain properties of the light $S$-, $P$- and $D$-wave resonances. Our goal is to investigate  sensitivity of these observables to the presence of an exotic meson and to guide future experimental analysis by identifying which combinations of  moments are most relevant for the identification 
of this resonance. In order to illustrate the sensitivity of the moments to exotic contributions, we discuss production of resonant $S$-, $P$-, $D$-waves in the forward direction which are produced dominantly by natural exchanges in the $t$ channel~\cite{Irving:1977ea} \cite{AlGhoul:2017nbp}. 

We consider two cases. In one we use the complete wave set ($S$-, $P$- and $D$-waves) and in the other we remove the $P$-wave.  By comparing the moments obtained in these two cases we can assess sensitivity to the presence of the exotic meson. 

The photon beam asymmetry corresponds to the difference in cross section for beam polarized parallel and perpendicular to the reaction plane, spanned by the momenta of the beam and the recoiling proton. In production of meson pairs there is an additional dependence on the direction of the relative momentum between the two mesons.

It is thus possible to give different definitions of the photon polarization asymmetry. 
Specifically, we consider the case when the decay angles are integrated over their whole domain, and when the relative momentum is fixed in the direction perpendicular to the reaction plane. 
We find that the maximal sensitivity of the beam asymmetry to the $P$-wave is obtained in the latter case. 
  
The paper is organized as follows. In Section~\ref{sec:model}, we describe the reaction model for the $\eta\pi^0$ photoproduction. In Section~\ref{sec:mom}, we calculate  moments of the di-meson angular distribution and in Section~\ref{sec:BA} we discuss the beam asymmetries. Our conclusions are presented in Section~\ref{sec:concl}. 

For clarity of presentation, all technical details are summarized in the Appendices. Specifically, in~\ref{sec:angle}, we describe the kinematics of $\eta\pi^0$ photoproduction and review the definition of the angular moments.
In~\ref{sec:linpol},  we derive formulas of the differential cross section in case of the linearly polarized beam. The relation between helicity amplitudes at high energy for a given naturality exchange are reviewed in~\ref{sec:parity}. In~\ref{sec:reflectivity} we extend the reflectivity basis to reactions with a photon beam.
Finally, the relations between the moments and the partial waves are summarized in~\ref{sec:app}.

\section{The model} \label{sec:model}

\begin{figure}[htb]
\begin{center}
\includegraphics[width=\linewidth]{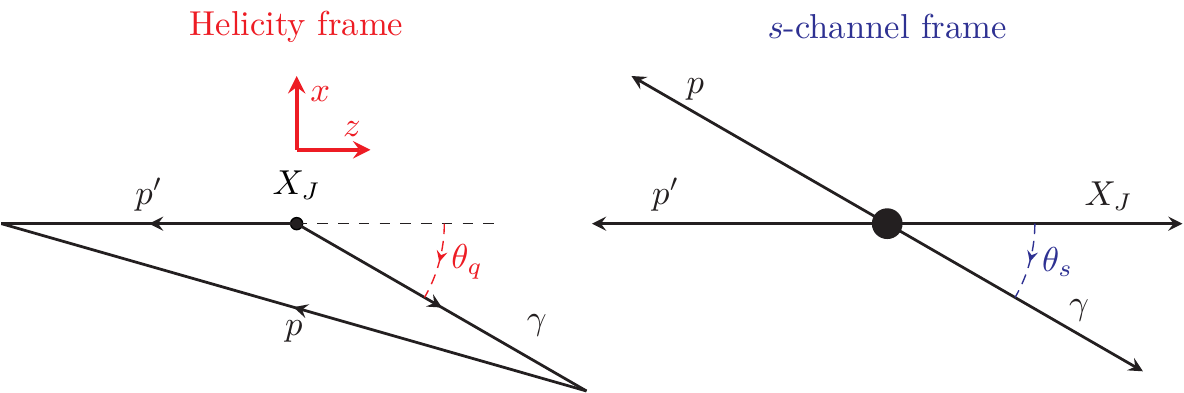}
\end{center}
\caption{\label{fig:schan}{\it Left:} the helicity frame, in which $X_J$, the $\eta\pi^0$ resonance of spin $J$, is at rest and quantized along the opposite direction of the recoil nucleon. {\it Right:} the $s$-channel frame, the center-of-mass frame of the reaction $\gamma p \to X_Jp$. 
The $s$-channel is obtained from the helicity frame by a boost along the $z$ axis. The boost leaves the helicity of $X_J$ unchanged.
The labels $\gamma$, $p$, and $p'$ stand for the beam, the 
nucleon target, and the recoiling nucleon, respectively.}
\end{figure}

We consider the reaction 
\begin{align} \label{eq:reaction2}
\vec\gamma (\lambda, p_\gamma)\ p(\lambda_1, p_N) \to \pi^0 (p_\pi)\ \eta (p_\eta)\ p (\lambda_2, p'_N).
\end{align} 
The helicities of the particles are defined in the
helicity frame,  the rest-frame of the $\eta\pi^0$ with the direction opposite to the recoil nucleon defining the $z$ axis
(see Fig.~\ref{fig:schan}). 
The amplitude for the reaction in  \eqref{eq:reaction2} is denoted by $A_{\lambda; \lambda_1\lambda_2} (\Omega)$, with $\Omega$ being 
the spherical angle determining the direction of the $\eta$ in this frame. 
The dependence on the remaining kinematical variables, \ie the total energy squared $s = (p_\gamma + p_N)^2$, the momentum transferred between the nucleons $t = (p_N-p'_N)^2$, and the $\eta\pi^0$ invariant mass squared $m_{\eta\pi^0}^2 = (p_\eta+p_\pi)^2$, is implicit. The direction of photon linear polarization is determined by the angle $\Phi$ which is measured with respect to the $\eta\pi$ production plane.
All the details and formulae are given in~\ref{sec:angle}. Below we summarize the key relations.  
In terms of the reaction amplitude $T$ the differential cross section is given by 
\begin{align} \nonumber
I(\Omega,\Phi) &\equiv \frac{\diff\sigma}{\diff t  \diff m_{\eta\pi^0} \diff\Omega\diff\Phi} \\
& = \kappa
\sum_{ \substack{\lambda, \lambda' \\ \lambda_1,\lambda_2}} A_{\lambda; \lambda_1\lambda_2} (\Omega) \rho^\gamma_{\lambda\lambda'}(\Phi) A_{\lambda'; \lambda_1\lambda_2}^* (\Omega),
\end{align}
with $\kappa$ containing all kinematical 
factors, {\it cf.} Eq.~\eqref{eq:pspace}. The photon spin density matrix $\rho^\gamma$ encodes the dependence on the polarization direction \cite{Schilling:1969um}. 
 Explicitly, 
\begin{align} \label{eq:IntPol}
I(\Omega,\Phi) & = I^0(\Omega) - P_\gamma I^1(\Omega) \cos 2 \Phi - P_\gamma I^2(\Omega) \sin 2 \Phi,
\end{align}
with $0<P_\gamma<1$, being the degree of linear polarization and 
\bsub \label{eq:int2}\begin{align}
I^0(\Omega) & = \frac{\kappa}{2}
\sum_{\lambda, \lambda_1,\lambda_2} A_{\lambda; \lambda_1\lambda_2} (\Omega) A_{\lambda; \lambda_1\lambda_2}^* (\Omega),\\
I^1(\Omega)& = \frac{\kappa}{2}
\sum_{\lambda, \lambda_1,\lambda_2} A_{-\lambda; \lambda_1\lambda_2} (\Omega) A_{\lambda; \lambda_1\lambda_2}^* (\Omega), \\
I^2(\Omega)& = i\frac{\kappa}{2}
\sum_{\lambda, \lambda_1,\lambda_2}\lambda A_{-\lambda; \lambda_1\lambda_2} (\Omega) A_{\lambda; \lambda_1\lambda_2}^* (\Omega).
\end{align} \esub
The partial wave amplitudes  $T^l$ are defined by 
\begin{align}
A_{\lambda; \lambda_1\lambda_2} (\Omega)  =  \sum_{\ell m} T^\ell_{\lambda m; \lambda_1\lambda_2} Y^{m}_\ell(\Omega).
\end{align}
Furthermore it is convenient to work in the so-called reflectivity basis which uses the following linear combination of the two, $\lambda_\gamma = \pm 1$ photon helicities  \begin{align}  \label{def:Teps2}
^{(\epsilon)}T^\ell_{m; \lambda_1\lambda_2} & 
\equiv   \frac{1}{2} \left[ T^\ell_{+1 m; \lambda_1\lambda_2}  - \epsilon \: (-1)^m T^\ell_{-1 -m; \lambda_1\lambda_2}\right],
\end{align}
with $m = -\ell, \cdots,  \ell$. 
As shown, in Appendix~\ref{sec:parity}, in the high-energy limit the amplitudes with  $\epsilon = +1(-1)$
are dominated by $t$-channel exchanges with naturality, $\eta=+1(-1)$, respectively.\footnote{The naturality is defined by $\eta = P(-1)^J$ for the exchange of spin $J$ and parity $P$. The reflectivity $\epsilon$ is the eigenvalue of the reflectivity operator, the symmetry through the reaction plane.}
Parity invariance implies
\begin{align} \label{eq:parity2}
^{(\epsilon)}T^\ell_{m; -\lambda_1-\lambda_2} & = \epsilon (-1)^{\lambda_1-\lambda_2}\  ^{(\epsilon)}T^\ell_{m; \lambda_1\lambda_2},
\end{align}
and we  take advantage of this constraint to define two sets of partial waves, 
\begin{align} \label{def:Leps2}
[\ell]^{(\epsilon)}_{m;0} &=\,  ^{(\epsilon)}T^\ell_{m; ++}, & 
[\ell]^{(\epsilon)}_{m;1} &=\, ^{(\epsilon)}T^\ell_{m; +-},
\end{align}
corresponding to nucleon helicity non-flip and flip, respectively. Here $[\ell] = S, P, D$ for $\ell = 0,1,2$ is the total spin of the $\eta\pi$ system. To summarize, in this basis for each $\ell$, there are $2  \times(2\ell+1)$ complex partial waves for nucleon helicity non-flip and independently $2  \times(2\ell+1)$ amplitudes describing  nucleon helicity flip. We note that in photoproduction, the reflectivity basis involves all values of $m$ while in the case of of spinless beams only the $m \ge 0$ spin projections enter~\cite{Chung:1974fq}. 

In the following we construct a model 
for $\eta\pi^0$ partial waves. Specifically, given the experimentally accessible mass range 
$m_{\eta\pi^0} < 2 \mbox{ GeV}$  we consider 
only the lowest three waves, $\ell =0,1, 2$~\cite{Adolph:2014rpp}. 
Moreover, we assume that the helicity-non-flip amplitudes dominate, and set the helicity-flip amplitudes to zero. This is not restrictive as the target is not polarized in GlueX, and the measured intensities are not sensitive to the details of the nucleon helicity structure. Finally, we consider only the amplitudes with $\epsilon = +$ based on the observation that natural parity exchanges are dominant in the energy range of interest~\cite{AlGhoul:2017nbp} \cite{Mathieu:2015eia}. 

The model is fully determined by the knowledge of the $2\ell+1$ projections of each spin $\ell$ wave. In order to reduce the number of projections, we can use the empirical observation of $s$-channel helicity conservation~\cite{Gilman:1970vi}\cite{Irving:1977ea}.\footnote{The $s$-channel is the center-of-mass frame of the reaction~\eqref{eq:reaction2}.} Fortunately, observables (moments and beam asymmetries) extracted in the helicity frame can be computed in the $s$-channel frame. 
As illustrated in Fig.~\ref{fig:schan}, the $s$-channel frame is related to the helicity frame by a boost along the $\eta\pi$ momenta. The boost leaves the helicities of the photon, of the $\eta\pi$ resonance and of the target proton invariant. On the contrary, the recoil proton helicity changes under this boost, but since this helicity are summed over when computing the moments and the beam asymmetries, the observables are invariant under this boost.  Consequently, the moments and the beam asymmetries in the $s$-channel frame and the helicity frame are identical. In the following, we take advantage of this equivalence and treat $m$ in $[\ell]_{m;k}^{(\epsilon)}$ as the spin projection of the $\eta\pi^0$ resonance of angular momentum $\ell$ in the $s$-channel frame.

The dominant $s$-channel helicity conserving amplitudes correspond to $m=1$. Therefore, requiring strict $s$-channel helicity conservation would remove the $S$-wave completely. We thus include the $m=0$ and $m=2$ contributions, which correspond to one unit of helicity flip at the photon vertex, and neglect the $m = -1$ and $m=-2$ projections.  Consequently, our basis is limited to the following waves
\begin{align} \label{eq:waves}
[\ell]^{(\epsilon)}_{m;k} =\left\{ S^{(+)}_{0}, P^{(+)}_{0,1}, D^{(+)}_{0,1,2} \right\}_{k=0}.
\end{align}

We now specify the dynamics of our model. We include the $a_0(980)$, $\pi_1(1600)$, $a_2(1320)$, and $a'_2(1700)$ resonances. We parameterize each resonance with a Breit-Wigner line shape,
\begin{align}
\Delta_R(m_{\eta\pi}) &=  \frac{m_R \Gamma_R}{m_R^2-m_{\eta\pi}^2 - im_R \Gamma_R}. 
\end{align}
$m_R$ and $\Gamma_R$ are the masses and total widths of the resonance $R$ respectively. 
For the $\pi_1(1600)$, $a_2(1320)$ and $a'_2(1700)$ resonances, we use the mass and width obtained from a recent fit to the $\pi^- p \to \eta^{(\prime)}\pi^- p$ COMPASS data~\cite{Rodas:2018owy}. For the $a_0$, we use the average mass and width quoted in the Review of Particle Physics~\cite{Tanabashi:2018oca}. The model parameters are summarized in Table~\ref{tab:bw}.

\begin{table}
\caption{Model parameters. 
The label $R$ stands for the resonance.
The mass ($m_R$) and width ($\Gamma_R$) of the resonances are given in GeV. The normalization ($N_R$) and the spin-flip coupling ($\delta_R$) are dimensionless.
\label{tab:bw}}
\centering{
\begin{tabular}{c|cc|cc|} 
$R$ & $m_R$ & $\Gamma_R$ & $N_R$ &$\delta_R$ \\[.1cm]
\hline
$a_0(980)$ & 0.980  &  0.075 & $\phantom{+}1.000$ & $\phantom{+}1.0$\\
$\pi_1(1600)$  &  1.564  & 0.492 & $-0.030$  &  $-5.0$\\
$a_2(1320)$ & 1.306  &  0.114 & $-0.109$ &   $-2.0$ \\
$a_2(1700)$ & 1.722  &  0.247 & $-0.036$ &   $-2.0$ 
\end{tabular}
}
\end{table}

We assume factorization of the production amplitude and include the high-energy limit of the angular momentum conservation factor $(\sqrt{-t})^{|m-1|}$ at the photon-resonance vertex. The contribution of the resonance $R$ to the wave $\ell$ reads:
\begin{align}\label{eq:model}
[\ell]^{(+)}_{m;0} &= N_0 N_R  \left(\delta_{R} \frac{\sqrt{-t}}{m_{R_{}}}\right)^{|m-1|} \Delta_{R}(m_{\eta\pi}) P_V(s,t)~.
\end{align}
$N_0$ is an arbitrary overall normalization, while $N_R$ is the normalization of each resonance relative to the $a_0(980)$, and $\delta_R$ is the helicity-flip coupling. For the $S$-wave we set $N_{a_0} = \delta_{a_0} = 1$. The remaining parameters $N_R$ and $\delta_R$ for the $P$- and $D$-waves in Eq.~\eqref{eq:model} are chosen to roughly reproduce the signs and the magnitude of the GlueX preliminary results~\cite{Gluexprem}.

The Regge propagator for the natural exchange takes the form
\begin{align}
P_V(s,t) &= \Gamma[1-\alpha(t)] \left( 1- e^{-i\pi \alpha(t)} \right) s^{\alpha(t)},
\end{align}
with $\alpha(t) = 0.5+0.9 t$, and with $s$ and $t$ expressed in GeV$^2$ in $P_V(s,t)$. 
The moments are calculated at $s = m_p^2 + 2 m_p E_\gamma$ with $E_\gamma = 9$ GeV and are integrated in the whole $t$ range. The Regge factor $P_V(s,t)$ provides an exponential suppression at large $|t|$. Since this factor is common to all waves, it contributes to the overall normalization for fixed $t$. The only $t$ dependence not common to all waves is due to the barrier factor $(\sqrt{-t})^{|m-1|}$.

\section{The moments} \label{sec:mom}
\begin{figure*}[htb]
\begin{center}
\includegraphics[width=0.95\linewidth]{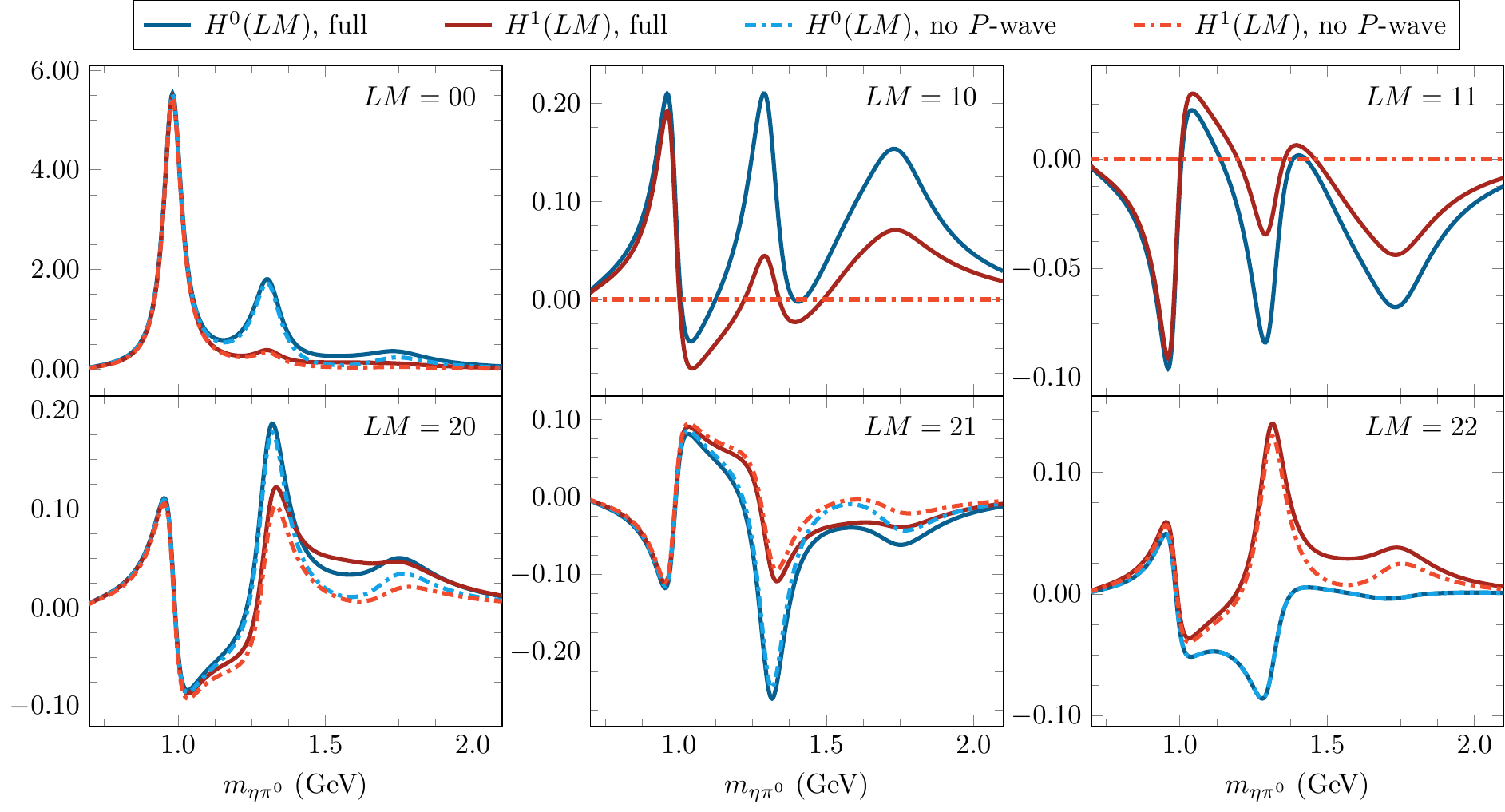}
\end{center}
\caption{Unpolarized $H^0(LM)$ moments (blue lines) compared to the polarized $H^1(LM)$ moments (red lines) for $L=0,1,2$, in the helicty frame, calculated with the models described in the text. The solid lines represent the complete model and the dashed lines the model without the $P$-wave. The moments are evaluated at $E_\gamma = 9$ GeV and integrated in $t$. \label{fig:mom1}}
\end{figure*}

\begin{figure*}[htb]
\begin{center}
\includegraphics[width=0.95\linewidth]{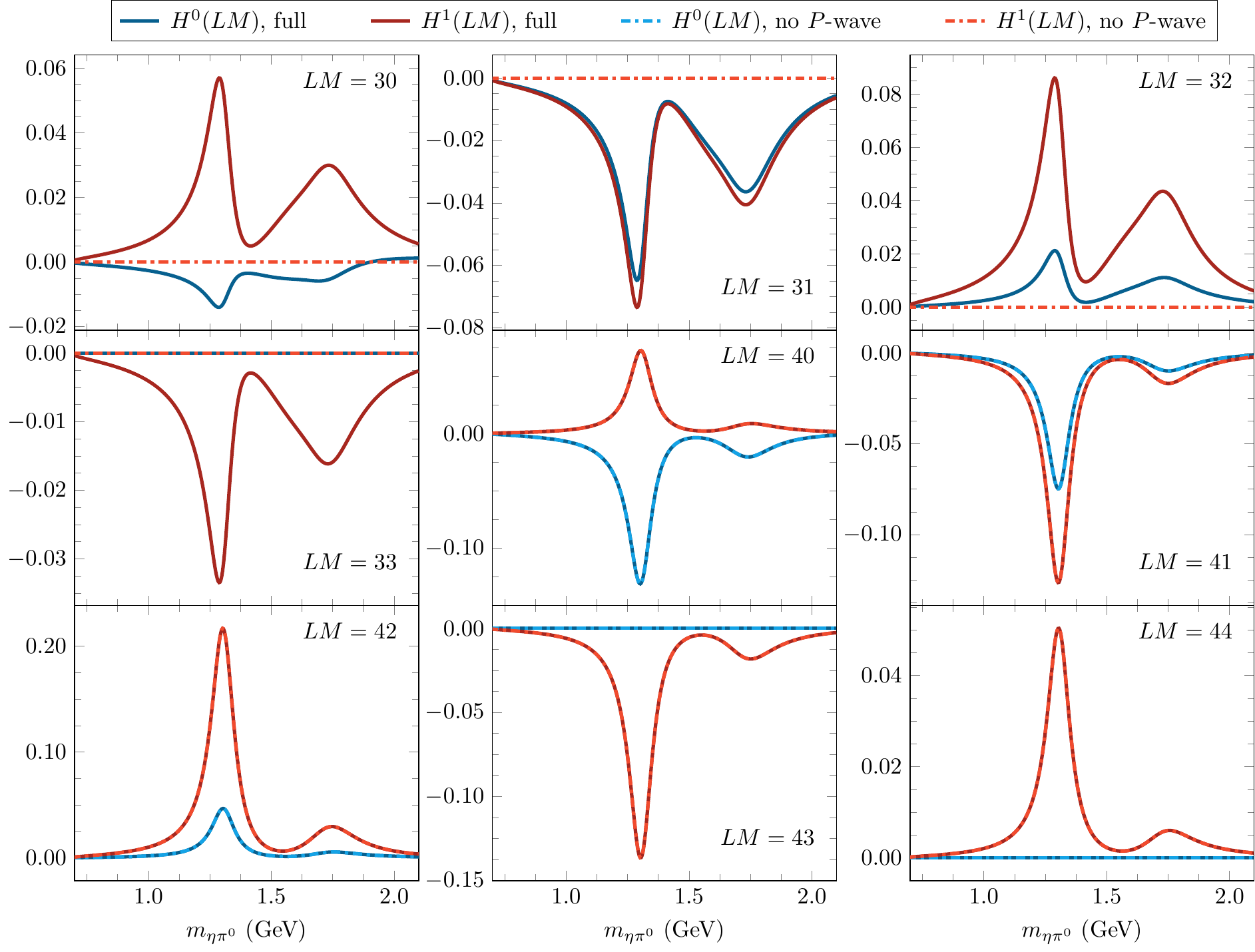}
\end{center}
\caption{Unpolarized $H^0(LM)$ moments (blue lines) compared to the polarized $H^1(LM)$ moments (red lines) for $L=3,4$ calculated, in the helicity frame, with the models described in the text. The solid lines represent the complete model and the dashed lines the model without the $P$-wave. The $L=3$ moments derived from the model without the $P$-wave are zero. The $L=4$ moments depend only on the $D$-wave and are therefore identical in both models, \ie, no sign of the exotic $P$-wave is to be expected in those moments. The moments are evaluated at $E_\gamma = 9$ GeV and are integrated in $t$. \label{fig:mom2}}
\end{figure*}

From the intensities in Eqs.~\eqref{eq:int2}, one computes the moments  
\begin{align} 
\nonumber
H^0(LM)  &= \frac{P_\gamma}{2} \int_\circ I(\Omega,\Phi) \, d^L_{M0}(\theta) \cos M\phi,\\
\nonumber
H^1(LM)  &= \int_\circ  I(\Omega,\Phi) \, d^L_{M0}(\theta) \cos M\phi\, \cos 2 \Phi ,\\
\im H^2(LM) & =- \int_\circ I(\Omega,\Phi)\, d^L_{M0}(\theta) \sin M\phi \, \sin 2 \Phi,
 \label{eq:int2mom2}
\end{align}
with $\int_\circ = (1/\pi P_\gamma)\int_0^\pi \sin\theta \diff \theta \int_0^{2\pi} \diff \phi \int_0^{2\pi} \diff \Phi$. Using the wave set in~\eqref{eq:waves},
one can extract the moments up to $L = 4$. In addition, since there are only waves with positive $m$ components (proved in Appendix~\ref{sec:reflectivity}) the moments fulfill the following relation
\begin{align}\label{eq:H2H1}
\im H^2(LM) & = -H^1(LM), & \text{for } M \geqslant 1.
\end{align}
Therefore, we only consider the moments $H^0(LM)$ and $H^1(LM)$ with $0\leqslant  L \leqslant4$ and $0\leqslant M \leqslant L$. The relations between the relevant moments and the partial waves restricted to the set~\eqref{eq:waves} are provided in~\ref{sec:app}. The relations~\eqref{eq:appmom1} show that it is advantageous to compare $H^1(LM)$ to $H^0(LM)$. Indeed, the difference $H^0(LM)-H^1(LM)$ is, in many cases, proportional to small partial wave interferences. Accordingly, the moments $H^0(LM)$ and $H^1(LM)$ for $L= 0$, $1$, and $2$ are shown in Fig.~\ref{fig:mom1}, and those for $L= 3$ and $4$ in Fig.~\ref{fig:mom2}. On both figures, the moments are computed with the $S$-, $P$- and $D$-waves but also with without the $P$-wave. The difference between the two models displays the sensitivity of the observables to the exotic wave.

Let us make some observations on Figs.~\ref{fig:mom1} and~\ref{fig:mom2}. From Eq.~\eqref{eq:appmom1}, we deduce the relation
\begin{align}
0 \leqslant H^1(00) \leqslant H^0(00).
\end{align}
It is worth pointing out that although the condition $0  \leqslant H^0(00)$ is always true  since $H^0(00)$ is proportional to the unpolarized cross section, the condition  $0 \leqslant H^1(00)$ is valid only in the absence of negative reflectivity components. 

The difference $H^0(00) - H^1(00)$, being proportional to the $m \neq 0$ components, vanishes when the $S$-wave dominates. From Fig.~\ref{fig:mom1}, we see that the $S$-wave describe the region $m_{\eta\pi^0} \leqslant 1.1$~GeV as expected from the resonance content of the model.

The strong $a_2(1320)$ peak in $H^{0}(00)$ is created by the dominance of the $m=1$ component of the $D$-waves. The components $m\neq 1$ are suppressed by the kinematical factor $(\sqrt{-t}/m_{a_2})^{|m-1|}$.
Let us also remark that $H^1(00)$ is proportional to the magnitude of the $m=0$ components.

Interestingly, we note from Eqs.~\eqref{eq:appmom1} that the difference $H^0(L1) -H^1(L1)$ with $L \leqslant 4$ is proportional to the $D^{(+)}_{2}$ wave. For instance, the moments $H^0(31)$ and $H^1(31)$ are very close since their difference is proportional to the interference of small waves $\re\left(P^{(+)}_{1} D^{(+)*}_{2}\right)$. In addition, the magnitude of the wave $D^{(+)}_{2}$ is directly measurable from the moment \mbox{$H^1(44) \propto |D^{(+)}_{2}|^2$} and its interference with the $S^{(+)}_{0}$, $P^{(+)}_{0}$, $D^{(+)}_{0}$, $P^{(+)}_{1}$ and $D^{(+)}_{1}$ waves are accessible with the moments  $H^0(22)$, $H^0(32)$, $H^0(42)$, $H^1(33)$ and $H^1(43)$ respectively. From Eqs.~\eqref{eq:appmom1}, we deduce the following relations between moments:
\bsub \label{eq:relH}
\begin{align}
H^0(11) - H^1(11) & = \frac{7}{15} \sqrt{10} H^1(33) \\
H^0(21) - H^1(21) & = 3\sqrt{\frac{6}{35}} H^1(43) \\
H^0(31) - H^1(31) & =  -\frac{1}{\sqrt{15}} H^1(33)
\end{align} \esub
Experimental deviations from these relations would imply that additional waves not included in the set~\eqref{eq:waves} are needed to describe the $\eta\pi^0$ system.

The presence of a $P$-wave is not clearly apparent in the leading moment $H^0(00)$, nor in any even moments. However, the odd $L$ moments are proportional to the interference between the $P$-wave and the $S$- and $D$-wave since $L+\ell+\ell'$ must be even in the sum~\eqref{eq:momsum}. Non-zero odd $L$ moments thus indicate the presence of the exotic wave. Interestingly, we note that the $a_2'(1700)$ is also more apparent in odd moments due to its interference with the $\pi_1(1600)$.


The observation of $P$-wave in odd moments can still be checked with the even moments.
In the case the $\eta\pi^0$ system is described with the waves in Eq.~\eqref{eq:waves}, it is straightforward to isolate the amplitude $|P_1^{(+)}|^2$ with specific linear combinations of even moments. With the definition $\Delta(LM) = H^0(LM)- H^1(LM)$, we obtain
\begin{subequations}\label{eq:relPp1}\begin{align}
|P^{(+)}_1|^2 
& = \frac{1}{2} \Delta(00) + \frac{21}{8} \Delta(40) + \frac{3}{4}\sqrt{\frac{35}{2}}  \Delta(44) \\
& = -\frac{5}{\sqrt{6}}  \Delta(22)  + \frac{15}{8} \Delta(40)  + \frac{3}{4}\sqrt{\frac{5}{14}}  \Delta(44) \\
& = -\frac{5}{2}  \Delta(20)  - \frac{15}{8}  \Delta(40)  + \frac{3}{4}\sqrt{\frac{35}{2}}  \Delta(44) \\
& = -\frac{5}{18}  \Delta(00)  - \frac{35}{36}  \Delta(20)  - \frac{35}{6\sqrt{6}}  \Delta(22) ~.
\end{align}\end{subequations}
If more waves than those in Eq.~\eqref{eq:waves} are needed to describe the system, then the linear combinations above would receive contributions from $F$- and higher waves. 
The first three relations are linearly independent and can be used to address systematic uncertainty related to the extraction of the moments. The fourth relation is a linear combination of the ones above, which however, can be convenient to use as it does not contain moments higher than $L = 2$.

From our moments analysis we can conclude that polarized moments $H^{1,2}(LM)$ provide additional constrains allowing to better identify the wave content of the $\eta\pi^0$ system. In particular, we have seen that the restriction $m \geqslant 0$ implies relations between moments that be checked experimentally. Moreover, the presence of an exotic wave could be directly identified from its interference with even waves in odd moments.

\section{Beam asymmetry} \label{sec:BA}
\subsection{General definition}
The beam asymmetry is defined as the difference in the intensity between polarization parallel $\Phi = 0$ and perpendicular $\Phi = \frac{\pi}{2}$ to the reaction plane, normalized to their sum. When two mesons are produced, the decay angles of one of the meson $\Omega = (\theta,\phi)$ have to be specified. A general definition of the beam asymmetry is thus
\begin{align} \label{eq:defBA}
\Sigma_{\cal D} &= \frac{1}{P_\gamma} \frac{\int_{\cal D} \left[I(\Omega,0) -  I(\Omega,\frac{\pi}{2}) \right] \diff\Omega}{\int_{\cal D} \left[I(\Omega,0) +  I(\Omega,\frac{\pi}{2}) \right] \diff\Omega}~.
\end{align}
In Eq.~\eqref{eq:defBA}, $\cal D$ is the domain of integration of the angular variables.
The subscript $\cal D$ indicates the dependence of the domain of integration in the definition of the beam asymmetry $\Sigma_{\cal D}$. 

\subsection{\boldmath $4\pi$ integrated beam asymmetry}
A standard choice is to integrate over the full kinematical range $\cos \theta \in [-1,1]$ and $\phi \in [0, 2\pi[$, or in short ${\cal D} = 4\pi$. The $4\pi$-integrated beam asymmetry $\Sigma_{4\pi}$ can equivalently be defined by
 \begin{align} \label{eq:defBA4pi}
\int_{4\pi} I(\Omega,\Phi) \diff \Omega & =  \sigma^0 \left(1 + P_\gamma  \Sigma_{4\pi} \cos 2 \Phi \right)~,
\end{align}
where the unpolarized integrated cross section is $\sigma^0 = H^0(00)$. 
Note that the term proportional to $\sin(2\Phi)$ in Eq.~\eqref{eq:IntPol} vanishes under the integration in Eq.~\eqref{eq:defBA4pi}. The sign in front of $P_\gamma \Sigma_{4\pi}$ is consistent with Eq.~\eqref{eq:defBA} and is such that natural (unnatural) exchanges contribute positively (negatively) to the beam asymmetry. This convention matches the convention of the CBELSA/TAPS collaboration, who extracted the $\eta\pi^0$ beam asymmetry for photon energies between $970$ MeV and $1650$ MeV~\cite{Gutz:2008zz}. The $\eta\pi^0$ beam asymmetry $\Sigma_{4\pi}$ has also been measured by the GRAAL experiment up to 1500 MeV~\cite{Ajaka:2008zz} and compared to the theoretical prediction based on the chiral unitary framework of Ref.~\cite{Doring:2010fw}. The definition in Eq.~\eqref{eq:defBA4pi} is similar to the one used in single pseudoscalar photoproduction~\cite{BDS1975}\cite{AlGhoul:2017nbp}, with the exception of the sign difference in front of $P_\gamma \Sigma_{4\pi}$. The  latter keeps the natural {\it vs.} unnatural exchange interpretation. The additional sign in single pseudoscalar photoproduction originates from the odd number of pseudoscalars in the final state.

The $4\pi$-integrated beam asymmetry can be extracted directly from the moments:
\begin{align} \label{eq:BA4toMom} 
\Sigma_{4\pi} &= \frac{-1}{P_\gamma} \frac{\int_{4\pi} I^1(\Omega)\diff \Omega}{\int_{4\pi}I^0(\Omega) \diff\Omega}  =\frac{H^1(00)}{H^0(00)}~.
\end{align}
%
As in the case of single pseudoscalar photoproduction, production mechanism {\it via} natural and unnatural exchanges contribute with opposite sign to $\Sigma_{4\pi}$. Explicitly, its expression in terms of partial waves reads
\begin{align} \label{eq:BA4pi_waves}
\Sigma_{4\pi} &= \frac{\sum_{k,\ell,m} (-1)^m \re\left([\ell]_{m;k}^{(+)} [\ell]_{-m;k}^{(+)*} - 
[\ell]_{m;k}^{(-)} [\ell]_{-m;k}^{(-)*}\right)}{\sum_{k,\ell,m} \left( \left|[\ell]_{m;k}^{(+)}\right|^2 + \left|[\ell]_{m;k}^{(-)}\right|^2 \right) }~.
\end{align}

Eq.~\eqref{eq:BA4pi_waves} can be understood as follows. The beam asymmetry represents the effect of the reflectivity operator, the reflection through the reaction plane. 
By construction, the partial waves in the reflectivity basis are invariant by reflection with $\epsilon$ being the eigenvalue of this operator. However the decay function $Y_\ell^m(\Omega)$ is in general not invariant and undergoes the change $Y_\ell^m(\Omega) \to (-1)^m Y_\ell^{-m}(\Omega)$ under reflection. Therefore only the combinations $\frac{1}{\sqrt{2}}\left([\ell]^{(\epsilon)}_{m;k} \pm (-1)^m [\ell]^{(\epsilon)}_{-m;k} \right) Y_\ell^m(\Omega)$ are invariant under reflection with the eigenvalue $\pm \epsilon$. The integration over the decay angles suppresses the interference between waves with different angular momenta by orthogonality of the $Y_\ell^m(\Omega)$, and the numerator of $\Sigma_{4\pi}$ is thus simply the difference
\begin{align} \nonumber
\sigma_0 \Sigma_{4\pi}  = \frac{\kappa}{2} \sum_{\epsilon,k,\ell,m} &\epsilon \bigg[ \left|[\ell]^{(\epsilon)}_{m;k} + (-1)^m [\ell]^{(\epsilon)}_{-m;k}\right|^2 \\
& - \left|[\ell]^{(\epsilon)}_{m;k} - (-1)^m [\ell]^{(\epsilon)}_{-m;k} \right|^2\bigg]~.
\label{eq:sigma2}
\end{align}
From Eq.~\eqref{eq:sigma2}, it is straightforward to find the range $-1 \leqslant \Sigma_{4\pi} \leqslant 1$. 
 
\subsection{\boldmath Beam asymmetry along the $y$ axis}
The beam asymmetry in which the two meson momenta were perpendicular to the reaction plane was introduced in Ref.~\cite{Criegee:1969qg}. With one of the mesons momentum having the angle $\Omega_y = (\frac{\pi}{2}, \frac{\pi}{2})$ along the $y$ axis, the definition of the beam asymmetry in Eq.~\eqref{eq:defBA} reduces to
\begin{align}\label{eq:defBAy}
\Sigma_y & = \frac{1}{P_\gamma} \frac{I(\Omega_y,0) - I(\Omega_y,\frac{\pi}{2})}{I(\Omega_y,0) + I(\Omega_y,\frac{\pi}{2})} = - \frac{I^1(\Omega_y)}{I^0(\Omega_y)}~.
\end{align}
The expression of intensities $I^\alpha(\Omega_y)$ with $\alpha = 0,1$ in terms of moments, truncated to $L=4$, is
\begin{align} \nonumber
\pm 4\pi I^\alpha(\Omega_y) & = H^\alpha(00) - \frac{5}{2} H^\alpha(20) - 5 \sqrt{\frac{3}{2}} H^\alpha(22) \\
& + \frac{27}{8} H^\alpha(40) + \frac{9}{2}\sqrt{\frac{5}{2}} H^\alpha(42) + \frac{9}{4}\sqrt{\frac{35}{2}}  H^\alpha(44)  
\label{eq:momy}
\end{align}

It was shown in the appendix of Ref.~\cite{Ballam:1971yd} that this definition leads to $\Sigma_y = \pm 1$ where a $\rho$ meson is produced {\it via} only natural or only unnatural exchanges in the process $\vec\gamma p \to \pi\pi p$.\footnote{It is worth noting that the convention adopted in Refs.~\cite{Criegee:1969qg} \cite{Schilling:1969um} \cite{Ballam:1971yd} differs by a minus sign from the definition~\eqref{eq:defBAy} since they focused only on the $P$-wave decay $\rho \to \pi\pi$. They sign was consistent with a beam asymmetry $\Sigma_y=1$ for a $P$-wave produced by naturality exchange, \cf  Eq.~\eqref{eq:BAy_L_nat}. } We will now  derive expression for $\Sigma_y$ when more than one wave populates the two mesons system. 

When the meson momenta are aligned with the $y$ axis, it is clear that the reflection through the reaction plane is equivalent to the parity transformation on the decay function $Y_\ell^m(\Omega_y) \to (-1)^\ell Y_\ell^m(\Omega_y)$. From this observation, we directly deduce that the results of the beam asymmetry along the $y$ axis for a system composed with a single wave $[\ell]^{\epsilon}_{m;k}$ is 
\begin{align} \label{eq:BAy_L_nat}
\Sigma_y  = \epsilon (-1)^\ell~,
\end{align}
since $[\ell]^{\epsilon}_{m;k} Y_\ell^m(\Omega_y)$ is invariant by reflection with the eigenvalue $\epsilon (-1)^\ell$. 

We can generalize this statement when the system is described by multiple waves by starting with the definition of the intensities
\begin{align}\label{eq:int0}
I^{\alpha}(\Omega_y) & = \sum_{\ell,\ell'} \sum_{m,m'} \rho^{\alpha, \ell\ell'}_{mm'} Y_{\ell}^m(\Omega_y)  Y_{\ell'}^{m'*}(\Omega_y)~.
\end{align}
We then note that $Y_{\ell}^{-m}(\Omega_y) = Y_{\ell}^m(\Omega_y)$. Moreover $Y_{\ell}^{m}(\Omega_y)\neq 0$ only when $m$ and $\ell$ have the same parity, {\it i. e.} $(-1)^m = (-1)^\ell$.\footnote{For completeness, we mention that $Y_{\ell}^{m}(\Omega_y) = i^\ell \sqrt{\frac{2\ell+1}{4\pi}} \sqrt{ \frac{(\ell-m)!}{(\ell+m)!}} \frac{(\ell+m-1)!!}{(\ell-m)!!}$,  $ \ell+m$ being  even.}
Using the parity relation~\eqref{eq:parity_rho}, we can re-write the intensities with $\alpha = 0,1$ as
\begin{align} \label{eq:int1}
I^{\alpha}(\Omega_y) & = \sum_{\ell,\ell'} \sum_{m,m'} (-1)^{m-m'} \rho^{\alpha, \ell\ell'}_{mm'} Y_{\ell}^m(\Omega_y)  Y_{\ell'}^{m'*}(\Omega_y)~.
\end{align}
Comparing Eqs.~\eqref{eq:int0} and~\eqref{eq:int1}, we see that the summation is restricted to $m$, $m'$, $\ell$ and $\ell'$ having the same parity. These restrictions and the relations~\eqref{eq:rho-pw} lead to the results:
\bsub \label{eq:I0I1}
\begin{align}
I^0(\Omega_y) & =2 \kappa \sum_{\epsilon, k,\ell, \ell'} \sum_{m, m'} [\ell]^{(\epsilon)}_{m;k} [\ell']^{(\epsilon)*}_{m';k} Y_\ell^m(\Omega_y) Y_{\ell'}^{m'*}(\Omega_y), \\
\nonumber
I^1(\Omega_y) & = -2 \kappa \sum_{\epsilon, k, \ell, \ell'} \epsilon (-1)^\ell \\
& \times \sum_{m,m'} [\ell]^{(\epsilon)}_{m;k}  [\ell']^{(\epsilon)*}_{m';k} Y_\ell^m(\Omega_y)Y_{\ell'}^{m'*}(\Omega_y),
\end{align}
\esub
where the summations are restricted to values of $\ell$, $\ell'$, $m$ and $m'$ having the same parity. 
From Eqs.~\eqref{eq:I0I1}, we see that $0 \leqslant |I^1(\Omega_y)| \leqslant I^0(\Omega_y)$ which yields $-1 \leqslant \Sigma_y \leqslant 1$.

At high energies, natural exchanges contribute only to waves with positive reflectivity, $\epsilon = +$, as demonstrated in Appendices~\ref{sec:parity} and~\ref{sec:reflectivity}. At GlueX, natural exchanges are expected to dominate~\cite{AlGhoul:2017nbp}. In the scenario where only natural exchanges contribute to the production of the $\eta\pi^0$, the beam asymmetry along the $y$ axis is $\Sigma_y \simeq (-1)^\ell$ in the mass region where the wave of spin $\ell$ dominates.  $\Sigma_y$ thus changes sign where an exotic (odd spin) wave dominate. $\Sigma_y$ is thus an interesting observable directly sensitive to exotic waves production in $\eta\pi$ photoproduction.

\subsection{Illustration of beam asymmetries}

\begin{figure}[t]
\begin{center}
\includegraphics[width=0.85\linewidth]{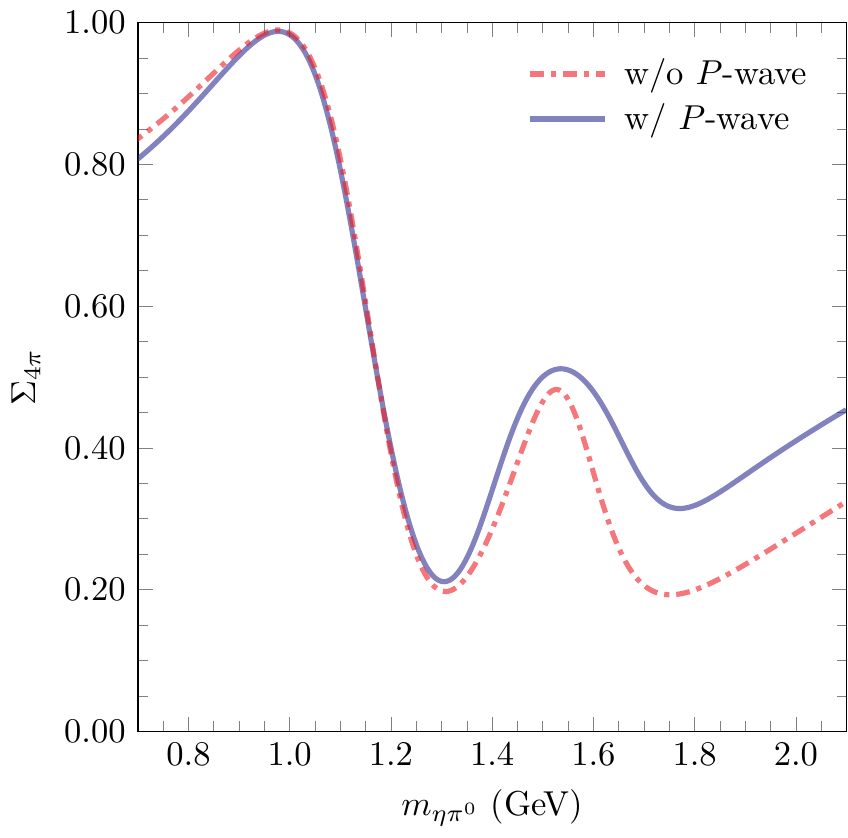}
\end{center}
\caption{Beam asymmetry $\Sigma_{4\pi}$ evaluated with the model described in the text at $E_\gamma = 9$ GeV and integrated in $t$. The solid blue line represents the complete model described in Sect.~\ref{sec:model}. The dashed-dotted red line is the model without the exotic $P$-wave.
\label{fig:BA4pi}}
\end{figure}

In this section we illustrate the differences between the beam asymmetries $\Sigma_{4\pi}$ and $\Sigma_y$ using our model described in Sect.~\ref{sec:model}.
To observe the impact of an exotic wave on the beam asymmetry,
we compare results in the complete model with the one without the $P$-wave.

In terms of our wave set~\eqref{eq:waves}, the $4\pi$-integrated beam asymmetry reads:
\begin{align}
\Sigma_{4\pi} & = \frac{r}{1+r} &
r & = \frac{|S_{0}^{(+)}|^2  + |P_{0}^{(+)}|^2  + |D^{(+)}_{0}|^2}{|P_{1}^{(+)}|^2
+ |D_{1}^{(+)}|^2  + |D_{2}^{(+)}|^2 }
\end{align}
Our model including only positive reflectivity component, $\Sigma_{4\pi}$ is always positive.
The beam asymmetry is represented on Fig.~\ref{fig:BA4pi} for the model with and without the $P$-wave.
The intensity is integrated over $t$ between $t_\text{max}(m_{\eta\pi^0})$ and $t_\text{min}(m_{\eta\pi^0})$. The $t$-dependence doesn't cancel in the ratio of the beam asymmetry since the $t$-dependence depends on the $m$ projection. 

We observe on Fig.~\ref{fig:BA4pi} that the model without the $P$-wave leads to a $\Sigma_{4\pi}$ very similar to the complete model. The reason is that the impact of the small $m=0$ $P$-wave component is overcome by the other waves, both in the numerator and denominator. We can conclude that the observable $\Sigma_{4\pi}$ is not sensitive to small exotic waves.  
In the $\eta\pi^0$ mass region close to the $a_0(980)$ peak, where the $S$-wave dominates, \mbox{$\Sigma_{4\pi} \sim 1$}  due to the dominance of positive naturality exchanges in the production. 

\begin{figure}[t]
\begin{center}
\includegraphics[width=0.85\linewidth]{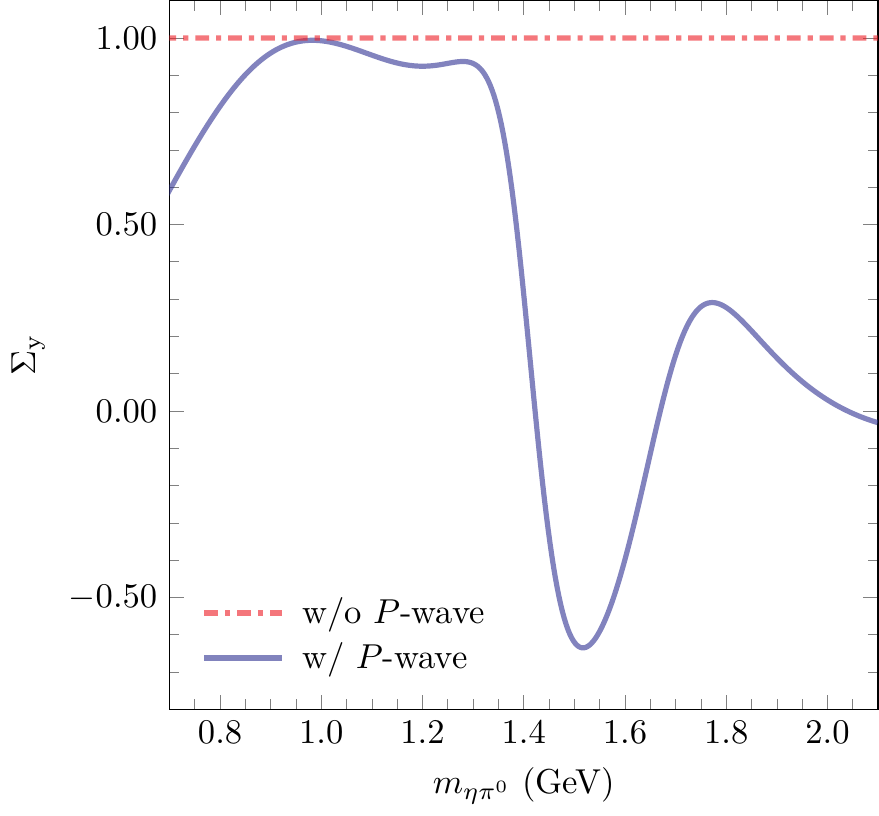}
\end{center}
\caption{Beam asymmetry $\Sigma_y$ evaluated with the model described in the text at $E_\gamma = 9$ GeV and integrated in $t$. The dashed-dotted red line is the model without the exotic $P$-wave. The presence of the $P$-wave around $m_{\eta\pi^0}\sim 1.5$ GeV is manifest in the full model.
\label{fig:BAy}}
\end{figure}

In terms of our waves, the beam asymmetry $\Sigma_y$ is given by:
\bsub\begin{align}
    \Sigma_y & = 1 - \frac{2|P_1^{(+)}|^2}{|P_1^{(+)}|^2 + \mathcal{R} } = \frac{\mathcal{R} - |P_1^{(+)}|^2}{\mathcal{R} + |P_1^{(+)}|^2}~,\\
\nonumber
    \mathcal{R} & = \frac{2}{3} |S_0^{(+)}|^2 + \frac{5}{6} |D_0^{(+)}|^2 + \frac{5}{4} |D_2^{(+)}|^2 
\\\nonumber  &
    - \frac{2\sqrt{5}}{3} \text{Re} \left( S_0^{(+)} {D_0^{(+)}}^\ast \right)
    - \sqrt{\frac{10}{3}} \text{Re} \left( S_0^{(+)} {D_2^{(+)}}^\ast \right) \\
&  + \frac{5}{\sqrt{6}}  \text{Re} \left( D_0^{(+)} {D_2^{(+)}}^\ast \right)~.
\end{align}\esub
The beam asymmetry along the $y$ axis, $\Sigma_y$, is illustrated on Fig.~\ref{fig:BAy}. As expected, the model without the $P$-waves leads to $\Sigma_y = 1$ in the whole range of $\eta\pi^0$ mass. However, $\Sigma_y$ computed with the complete model presents a significant depletion around $1.5$ GeV produced by the enhancement of the $P$-wave in this observable. The beam asymmetry does not reach $\Sigma_y = -1$ at the peak since the nearby $a_2(1320)$ and $a_2(1700)$ contribute to $\Sigma_y$ in the mass region of the $\pi_1(1600)$. However, although the small $\pi_1(1600)$ is not really apparent in the differential cross section, its effect is enhanced in $\Sigma_y$. The depletion produced by the odd wave is sharp and significant, suggesting that $\Sigma_y$ is an observable highly sensitive to exotic waves.

From an experimental point of view, the meson momentum is never exactly aligned with the $y$ axis. $\Sigma_y$ can be computed from the moments thanks to Eq.~\eqref{eq:momy}. Alternatively, $\Sigma_y$ can be approximated by the beam asymmetry binned around the $y$ axis. We will denote the quantity $\Sigma_{y\pm\tau}$, the beam asymmetry~\eqref{eq:defBA} with the integration domain $\theta \in [\frac{\pi}{2} - \tau, \frac{\pi}{2} + \tau]$ and $\phi \in [\frac{\pi}{2} - \tau, \frac{\pi}{2} + \tau]$. Let us point out that the properties of $\Sigma_y$ hold when the meson momenta are along the $y$ axis in either direction. In other words, one can experimentally measure $\Sigma_{y\pm\tau}$ by combining the data binned in $\phi \in [\frac{\pi}{2} - \tau, \frac{\pi}{2} + \tau] \cup  [\frac{3\pi}{2} - \tau, \frac{3\pi}{2} + \tau]$, and $\theta \in [\frac{\pi}{2} - \tau, \frac{\pi}{2} + \tau]$.

\begin{figure*}[t]
\begin{center}
\includegraphics[width=\linewidth]{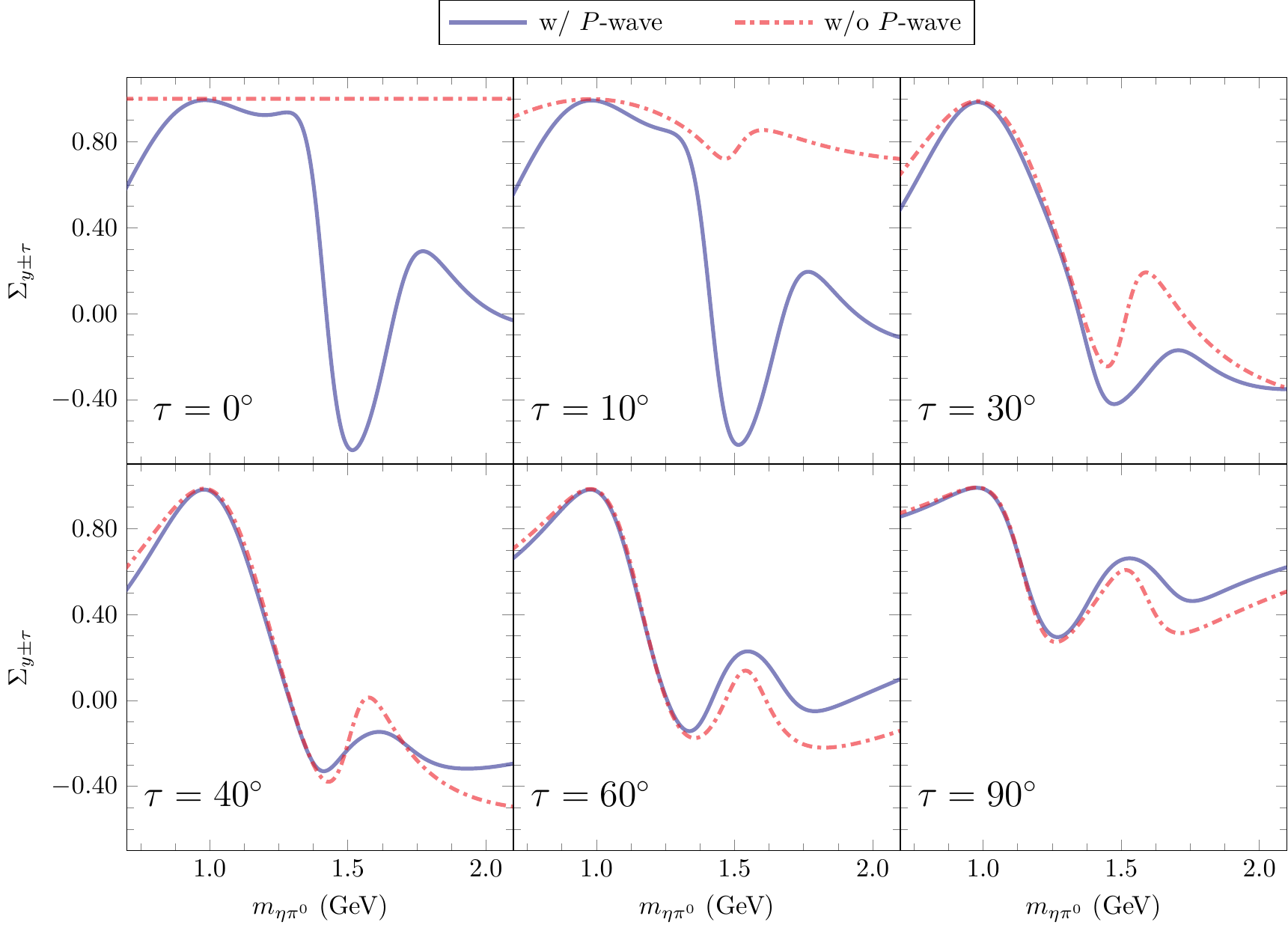}
\end{center}
\caption{Evolution of the beam asymmetry $\Sigma_{y\pm \tau}$ for $\tau$ between $0^\circ$ and $90^\circ$. The model including $S$-,$P$- and $D$-waves is shown in solid blue lines. The model including only $S$- and $D$-waves is shown in dashed-dotted red lines. 
The models are evaluated at $E_\gamma = 9$ GeV and integrated in $t$. \label{fig:BAy-2}}
\end{figure*}

As the opening angle $\tau$ increases $\Sigma_{y\pm \tau}$ should approach the $4\pi$-integrated beam asymmetry since $\Sigma_{y\pm 90^\circ} = \Sigma_{4\pi}$. Fig.~\ref{fig:BAy-2} illustrate how the observable $\Sigma_{y\pm \tau}$ varies as $\tau$ increases. $\Sigma_{y\pm \tau}$ is computed with our complete model and with the model without the $P$-wave. We note that the complete model is almost not sensitive to $\tau$ as long as $\tau \leqslant 10^\circ$. However the model featuring  only even waves displays a bigger sensitivity to $\tau$. The reason is that, without the $P$-wave, $\Sigma_{y\pm \tau}$ is the ratio of small intensities and both the numerator and denominator are sensitive to variation of the opening angle. {\it A contrario}, in the presence of a $P$-wave, both the numerator and denominator of $\Sigma_{y\pm \tau}$ are large and are less sensitive to variation in the parameter $\tau$. This conclusion is valid as long as the opening angle remains small. For larger values $\tau > 30^\circ$, the observable is no longer sensitive to the $P$-wave, as can be seen on Fig.~\ref{fig:BAy-2}. At this point, it is worth stressing that the asymmetry $\Sigma_y$ can also be computed from the measured intensities, Eq.~{\eqref{eq:defBAy}}.

\section{Conclusions} \label{sec:concl}

The paper presents a simple model to illustrate moments of the angular distribution of the $\eta\pi^0$ photoproduction with a linearly polarized beam. The model features $S$-, $P$-, $D$-waves produced by natural exchanges, whose parameters were guided by $s$-channel helicity conservation. The main motivation behind the $\eta\pi^0$ channel is the studies of exotic mesons, whose lightest candidate is expected in the $P$-wave. We showed that a non-zero $P$-wave would be directly observable from its interference with even waves in moments with odd angular momenta. It was also shown that some specific linear combination of moments, depending on the maximum angular momentum waves contributing to the $\eta\pi^0$ system, allow to isolate the $P$-wave.

For a given wave content, kinematical relations between the moments are derived. For instance, we demonstrated the relation $\im H^2(LM) =  -H^1(LM)$ for $M \geqslant 1$,  when the wave set contains only positive $m$ components. We demonstrated how the relations between the partial waves and the moments can be read out directly from the moments. By comparing the experimental moments with their expression in term of partial waves, it will be possible to deduce the dominant waves needed to describe the $\eta\pi^0$ system. 

Another set of observables currently under extraction by the GlueX collaboration are the beam asymmetries. We proposed a definition of the beam asymmetry, $\Sigma_{\cal D}$, in which the decay angles of the meson are integrated over a region $\cal D$ of the sphere. We show that when the decay angles are integrated over the whole sphere, the resulting beam asymmetry $\Sigma_{4\pi}$ is not very sensitive to the presence of a $P$-wave. However,  when the meson momenta are perpendicular to the reaction plane, the beam asymmetry, called $\Sigma_y$, is sensitive to the parity of the wave. In particular, in the mass region dominated by a wave of angular momentum $\ell$ produced by natural exchange, the beam asymmetry is $\Sigma_{y} = (-1)^\ell$, at high energy. We concluded that the beam asymmetry along the $y$ axis is an important observable in the search for exotic mesons with the GlueX experiment. Finally we tested the sensitivity of $\Sigma_{y\pm \tau}$, in which the decay angles are binned within a opening angle of $\tau$ around the $y$ axis. We showed that the model with and without the $P$-wave are clearly distinguishable with an opening angle up to $\tau = 10^\circ$. But for large opening angle $\tau > 30^\circ$, the beam asymmetry $\Sigma_{y\pm \tau}$ is no longer sensitive to the $P$-wave.

The illustration of the observables depends on the model presented in Sect.~\ref{sec:model}. The interested reader has the possibility to change the model parameters and the kinematical variables in the online version of the model~\cite{JPACweb} \cite{Mathieu:2016mcy}. The online version also offers the possibility to calculate the moments at a specific $t$, instead of integrating over $t$.

\acknowledgments
We thank A.~Austregesilo, S. Dobbs, D. Glazier, C.~Gleason, C.~Salgado, E.~Smith, 
J.~Stevens and A. Thiel for useful comments and discussions. 
V.M. acknowledges support from Comunidad Aut\'onoma de Madrid through 
Programa de Atracci\'on de Talento Investigador 2018 (Modalidad 1).
This work was supported by the U.S.~Department of Energy under Grants
No.~DE-AC05-06OR23177 
and No.~DE-FG02-87ER40365, 
the U.S.~National Science Foundation under Grant 
No.~PHY-1415459, 
by the Ministerio de Ciencia, Innovaci\'on y Universidades (Spain) under Grants No.~FPA2016-77313-P and No.~FPA2016-75654-C2-2-P, 
by  PAPIIT-DGAPA (UNAM, Mexico) under Grant No.~IA101819, 
and CONACYT (Mexico) under Grants No.~251817  
and No.~A1-S-21389. 

\appendix

\section{Angular distributions} \label{sec:angle}
\begin{figure}[htb]
\begin{center}
\includegraphics[width=\linewidth]{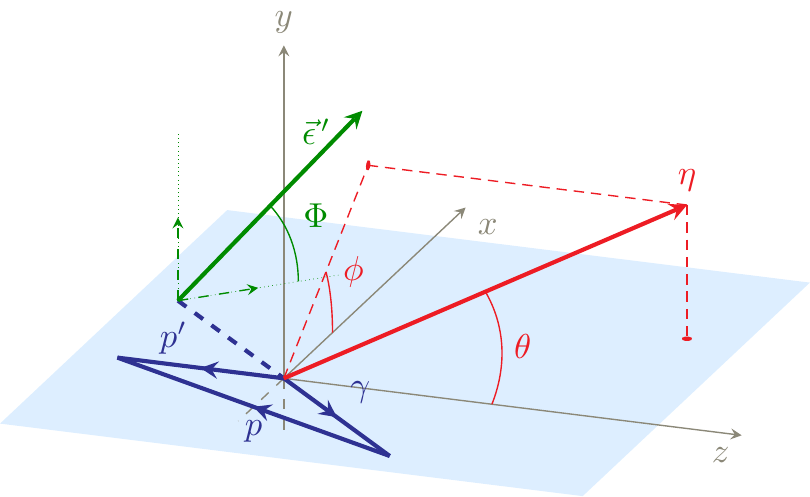}
\end{center}
\caption{\label{fig:frame}Definition of the angles in the helicity frame. The reaction plane $xz$, containing the momenta of the photon beam ($\gamma$), the nucleon target ($p$) and recoiling nucleon ($p'$), is in blue. 
$\theta$ and $\phi$ are the polar and azimuthal angles of the $\eta$. The polarization vector of the photon forms an angle $\Phi$ with the reaction plane.}
\end{figure}

We consider the reaction 
\begin{align} \label{eq:reaction}
\vec\gamma (\lambda, p_\gamma)\ p(\lambda_1, p_N) \to \pi^0 (p_\pi)\ \eta (p_\eta)\ p (\lambda_2, p'_N).
\end{align} 
The photon beam is linearly polarized with an angle $\Phi$ with respect to the reaction plane $xz$, the plane formed by the beam, the target and the recoiling nucleon in the center of mass of the $\eta\pi$ system.  As illustrated on Fig.~\ref{fig:frame}, the $z$ axis is defined as the opposite direction of the recoiling nucleon. The normal to the reaction plane is $\bm y = \bm p'_N \times \bm p_\gamma / |\bm p'_N \times \bm p_\gamma|$ and the $x$ axis is given by right-hand rule, $\bm x  = \bm y \times \bm z$.\footnote{We use the boldface font to indicate spatial three-vectors.} With this choice of axes, $\Omega = (\theta,\phi)$ are the angles of the $\eta$. This convention for the axes corresponds to the helicity frame.  In Eq.~\eqref{eq:reaction}, $\lambda$, $\lambda_1$ and $\lambda_2$ are the helicities of the beam, target and recoiling nucleon, respectively. 

The Mandelstam variables are the total energy squared $s = (p_\gamma + p_N)^2$, the momentum transferred between the nucleons $t = (p_N-p'_N)^2$, and the $\eta\pi^0$ invariant mass squared $m_{\eta\pi^0}^2 = (p_\eta+p_\pi)^2$. The dependence in the Mandelstam variables $s$, $t$ and $m_{\eta\pi^0}$ will be implicit thorough the paper as we are mainly focusing on the angular dependence. The amplitude for the reaction \eqref{eq:reaction} is $A_{\lambda; \lambda_1\lambda_2} (\Omega)$.
The $\Phi$-dependence of the intensity is encoded in the density matrix of the photon $\rho^\gamma$ \cite{Schilling:1969um} and the differential cross section in photoproduction is, with the flux $F_I = 2 (s-m_N^2)$,
\begin{align} \nonumber
\diff\sigma & = (2\pi)^4 \delta^4 ( \Sigma\;  p) \frac{1}{F_I} \frac{1}{(2\pi)^9} \frac{\diff^3 \bm p_\pi}{2 E_\pi} 
\frac{\diff^3 \bm p_\eta}{2 E_\eta} \frac{\diff^3 \bm p_N}{2 E_N} \frac{1}{2} \\
& \times \sum_{ \substack{\lambda, \lambda' \\ \lambda_1,\lambda_2}} A_{\lambda; \lambda_1\lambda_2} (\Omega) \rho^\gamma_{\lambda\lambda'}(\Phi) A_{\lambda'; \lambda_1\lambda_2}^* (\Omega). 
\end{align} 
In the rest frame of $\eta\pi^0$, the measured intensity becomes
\begin{align} \nonumber
I(\Omega,\Phi)&=  \frac{\diff\sigma}{\diff t  \diff m_{\eta\pi^0} \diff\Omega\diff\Phi} \\
&= \kappa
\sum_{ \substack{\lambda, \lambda' \\ \lambda_1,\lambda_2}} A_{\lambda; \lambda_1\lambda_2} (\Omega) \rho^\gamma_{\lambda\lambda'}(\Phi) A_{\lambda'; \lambda_1\lambda_2}^* (\Omega). 
\end{align}
We include all numerical factors in the phase space factor ($m_x$ is the mass of particle $x$),\footnote{The phase space factor is often absorbed in a redefinition of the amplitudes $\widehat T \equiv \sqrt{\kappa} T$ since it is numerically more stable to extract $\widehat T$ from data near the $\eta\pi^0$ threshold, where $\kappa \to 0$.}
\begin{align} \label{eq:pspace}
\kappa & = \frac{1}{(2\pi)^3} \frac{1}{4\pi} \frac{1}{2\pi} \frac{\lambda^{1/2}(m_{\eta\pi^0}^2, m_\pi^2, m_\eta^2)}{16m_{\eta\pi^0}(s-m^2_N)^2}\frac{1}{2}.
\end{align}
The triangle function is $\lambda(a,b,c) = a^2+b^2+c^2-2(ab+bc+ca)$.

We next expand the amplitude in $\eta\pi^0$ partial waves:
\begin{align}
A_{\lambda; \lambda_1\lambda_2} (\Omega) = \sum_{\ell m} T^\ell_{\lambda m; \lambda_1\lambda_2} Y^{m}_\ell(\Omega).
\end{align}
We can further make the $\Phi$ dependence explicit by decomposing the spin density matrix of the photon. Using a matrix notation $\rho^\gamma_{\lambda \lambda'} \equiv (\rho_\gamma)_{\lambda \lambda'}$, we expand it in a base of Hermitian $2\times 2$ matrices composed of the unity matrix $I$ and the Pauli matrices $\bm \sigma$:
\begin{align}
\rho_\gamma(\Phi) = \frac{1}{2} I + \frac{1}{2} \bm P_\gamma(\Phi)\cdot \bm \sigma.
\end{align}
The vector $\bm P_\gamma$ encodes the information about the polarization of the beam~\cite{Schilling:1969um}. Similarly, one defines
\begin{align}
I(\Omega,\Phi) & = I^0(\Omega) +\bm I(\Omega)\cdot \bm P_\gamma(\Phi),
\end{align}
with the vector of polarized intensities $\bm I = (I^1,I^2,I^3)$.  
The angular distribution can be expanded in unpolarized moment $H^0$ and polarized moments $\bm H = (H^1,H^2,H^3)$ via
\begin{subequations}\label{eq:defHa}
\begin{align}
I^0 ( \Omega) & = \sum_{LM} \left(\frac{2L+1}{4\pi} \right) H^0 (LM) D^{L*}_{M0} (\phi, \theta,0), \\
\bm I ( \Omega) & = -\sum_{LM} \left(\frac{2L+1}{4\pi} \right) \bm H (LM) D^{L*}_{M0} (\phi, \theta,0).
\end{align}\end{subequations}
The extra minus sign in the definition of $\bm H$ ensures that $H^1(00)$ is positive for positive reflectivity waves, cf.~\ref{sec:reflectivity}. 
The moments are expressed in terms of the $\eta\pi^0$ SDME:
\begin{subequations}\label{eq:momsum}
\begin{align} 
H^0(LM) &=\sum_{\substack{\ell \ell' \\ m m'} } \left(  \frac{2\ell'+1}{2\ell+1} \right)^{1/2} 
C^{\ell 0}_{\ell' 0 L0} C^{\ell m}_{\ell'm' LM}\ \rho^{\alpha,\ell \ell'}_{mm'}, \\
\bm H(LM) &=  -\sum_{\substack{\ell \ell' \\ m m'} } \left(  \frac{2\ell'+1}{2\ell+1} \right)^{1/2} 
C^{\ell 0}_{\ell' 0 L0} C^{\ell m}_{\ell'm' LM}\ \bm \rho^{\ell \ell'}_{mm'}
\end{align}\end{subequations}
where the $C^{\ell 0}_{\ell' 0 L0}$ and $C^{\ell m}_{\ell'm' LM}$ are the Clebsch-Gordan coefficients. They impose that $L+ \ell+\ell'$ be an even integer and restrict the summation to $M+m' = m$. 
The spin density matrices $\rho^{\alpha, \ell\ell'}_{mm'} = (\rho^0, \bm \rho)^{\ell\ell'}_{mm'}$ are given by: 
\begin{equation}\label{eq:rho}
    \rho^{\alpha,\ell \ell'}_{mm'} = \frac{\kappa}{2} \sum_{\lambda, \lambda_1,\lambda_2} T^\ell_{\lambda m; \lambda_1\lambda_2} \sigma^\alpha_{\lambda\lambda'} T^{\ell' *}_{\lambda' m'; \lambda_1\lambda_2}~,
\end{equation}
with $\sigma^{\alpha} = (I,\bm{\sigma})$. 
More explicitly, the SDME read
\bsub\begin{align}
\rho^{0,\ell \ell'}_{mm'} & = \frac{\kappa}{2}\sum_{\lambda, \lambda_1,\lambda_2} T^\ell_{\lambda m; \lambda_1\lambda_2} T^{\ell' *}_{\lambda m'; \lambda_1\lambda_2}, \\
\rho^{1,\ell \ell'}_{mm'} & =  \frac{\kappa}{2}\sum_{\lambda,\lambda_1,\lambda_2} T^\ell_{-\lambda m; \lambda_1\lambda_2} T^{\ell' *}_{\lambda m'; \lambda_1\lambda_2}\\
\rho^{2,\ell \ell'}_{mm'} & =  i\frac{\kappa}{2}\sum_{\lambda,\lambda_1,\lambda_2} \lambda T^\ell_{-\lambda m; \lambda_1\lambda_2} T^{\ell' *}_{\lambda m'; \lambda_1\lambda_2}, \\
\rho^{3,\ell \ell'}_{mm'} & =  \frac{\kappa}{2}\sum_{\lambda,\lambda_1,\lambda_2} \lambda T^\ell_{\lambda m; \lambda_1\lambda_2} T^{\ell' *}_{\lambda m'; \lambda_1\lambda_2}
\end{align} \esub

The amplitudes $T^\ell_{\lambda m; \lambda_1 \lambda_2}$, and thus the SDME $\rho^{\alpha,\ell\ell'}_{mm'}$, depend on the frame. For completeness, we mention that the formalism of this section, although derived in the helicity frame, equally applies to any other $\eta\pi^0$ rest frame.  
In practice, the SDME are extracted experimentally in a $\eta\pi^0$ rest frame, either the GJ frame or the helicity frame and the theoretical models are built in either the $s$-channel or the $t$-channel frame.\footnote{The $s$-channel frame is the center-of-mass frame of the reaction $\gamma p \to \eta\pi^0 p$. The $t$-channel frame is the center-of-mass frame of the reaction $\bar p p \to \gamma \eta\pi^0$.}
The $s$-channel ($t$-channel) frame and the helicity (GJ) frame lead to the same SDME as demonstrated in the Appendix of Ref.~\cite{Mathieu:2018xyc}. The moments built in the $s$-channel can thus be compared to the ones extracted in the helicity frame. 
The relation between the the helicity and GJ frames is a rotation around the $y$ axis (with $\alpha = 0,1,2,3)$:
\label{eq:rot}
\begin{align} 
    \rho_{mm'}^{\alpha, \ell \ell'}|_{\text{GJ}} & = \sum_{\lambda \lambda'} d^\ell_{m \lambda}(\theta_q)\, \rho_{\lambda\lambda'}^{\alpha, \ell \ell'}|_{\text{hel}} \, d^{\ell'}_{m' \lambda'}(\theta_q), \\
    \left. H^\alpha(LM)\right|_\text{GJ} & = \sum_{M'} \left. H^\alpha(LM')\right|_\text{hel} d^L_{MM'}(\theta_q)
\end{align}
with $\cos \theta_q = (\beta-z_s)/(\beta z_s -1)$, $\beta = \lambda^{1/2}(s,m_N^2, m_{\eta\pi^0}^2) / (s-m_N^2 + m_{\eta\pi^0}^2)$ and $z_s = \cos\theta_s$ the cosine of the scattering angle between the target an recoiling nucleon in the center-of-mass frame. The angles $\theta_q$ and $\theta_s$ are indicated on Fig.~\ref{fig:schan}.

The spin density matrix is Hermitian $\left[\rho^{\alpha, \ell' \ell}_{m'm}\right]^* = \rho^{\alpha, \ell \ell'}_{mm'}$ and so \mbox{$\left[H^{\alpha}(LM)\right]^* =(-1)^M H^\alpha(L-M)$}. 
Under a parity transformation the decay angles transform as $(\theta,\phi) \to (\pi-\theta, \pi+\phi)$ which induces the transformation $Y^m_{\ell}(\Omega) \to (-1)^\ell Y^m_{\ell}(\Omega)$. 
Taking into account the intrinsic parity of the particles, the invariance under parity implies the relation (since $|\lambda|=1$)
\begin{align} \label{eq:parityT2}
T^\ell_{-\lambda -m; -\lambda_1-\lambda_2} &=  -(-1)^{m + \lambda_1-\lambda_2} T^\ell_{\lambda m; \lambda_1\lambda_2}~. 
\end{align}
The parity relations and the properties of the Clebsch-Gordan coefficients lead to the following relations for the SDME
\bsub \label{eq:parity_rho}\begin{align}
\rho^{0, \ell\ell'}_{mm'} & = \phantom{-}(-1)^{m-m'} \rho^{0, \ell\ell'}_{-m-m'}, \\ 
\rho^{1, \ell\ell'}_{mm'} & = \phantom{-}(-1)^{m-m'} \rho^{1, \ell\ell'}_{-m-m'}, \\ 
\rho^{2, \ell\ell'}_{mm'} & = -(-1)^{m-m'} \rho^{2, \ell\ell'}_{-m-m'}, \\
\rho^{3, \ell\ell'}_{mm'} & = -(-1)^{m-m'} \rho^{3, \ell\ell'}_{-m-m'},
\end{align}\esub
and similarly for the moments
\bsub\label{eq:parity_mom}\begin{align} 
H^{0}(LM) &=\phantom{-}(-1)^MH^{0}(L-M),\\ 
H^{1}(LM) &=\phantom{-}(-1)^MH^{1}(L-M),\\ 
H^{2}(LM) &= -(-1)^MH^{2}(L-M), \\
H^{3}(LM) &= -(-1)^MH^{3}(L-M).
\end{align}\esub
It follows that the moments $H^\alpha(LM)$ are purely real for $\alpha = 0,1$ and purely imaginary for $\alpha=2,3$.
Using these relations, one can write the intensity as
\begin{align} \nonumber
I^{0}(\Omega) & =  \sum_{L,M\geqslant 0} \left(\frac{2L+1}{4\pi} \right) \tau(M) H^{0}(LM) d^{L}_{M0} (\theta) \cos M\phi,  \\ \nonumber
I^{1} (\Omega)& = - \!\!\!\!\! \sum_{L,M\geqslant 0} \left(\frac{2L+1}{4\pi} \right) \tau(M) H^{1}(LM) d^{L}_{M0} (\theta) \cos M\phi, 
\\ \nonumber
I^2 (\Omega)& = 2 \sum_{L,M>0} \left(\frac{2L+1}{4\pi} \right)  \im H^{2}(LM) d^{L}_{M0} (\theta) \sin M\phi, \\
I^3 (\Omega)& = 2 \sum_{L,M>0} \left(\frac{2L+1}{4\pi} \right)  \im H^{3}(LM) d^{L}_{M0} (\theta) \sin M\phi,
\label{eqs:Ialpha}
\end{align}
with the definition $\tau(M) = (2-\delta_{M,0})$.

\section{Linearly polarized beam} \label{sec:linpol}
In this section we particularize our formulas for the case of a linearly polarized beam. In the GJ frame, the polarization vector of the photon is $\bm \varepsilon(\Phi) = (\cos\Phi, \sin \Phi,0)$, which leads to the pure photon state~\cite{Schilling:1969um}
\begin{equation}\label{eq:gammapure}
\ket{\Phi} = - \frac{1}{\sqrt{2}} \left[ e^{-i \Phi} \ket{+} - e^{i \Phi} \ket{-} \right].
\end{equation}
The helicity states $\ket{\pm} \equiv \ket{\lambda=\pm 1}$ are defined in the Cartesian basis by $\bm \varepsilon(\lambda=\pm 1) = (\mp 1, -i,0)/\sqrt{2}$~\cite{Walker:1968xu}. In the helicity frame, both $\bm \varepsilon(\lambda)$ and $\varepsilon(\Phi)$ rotates by $-\theta_q$ around the $y$ axis and thus Eq.~\eqref{eq:gammapure}, and all other equations in this Appendix remain unchanged in the helicity frame. 
The density matrix for the pure photon state in Eq.~\eqref{eq:gammapure} is thus
\begin{align}
\rho_{\gamma,\text{pure}}(\Phi) &= \ket{\Phi}\bra{\Phi} = \frac{1}{2} 
\begin{pmatrix} 1 & - e^{-2i \Phi} \\ - e^{2i \Phi} & 1\end{pmatrix}.
\end{align}
To describe a partially linearly polarized beam we consider a statistical mixture of the pure states $\ket{\pm}$ and $\ket{\Phi}$. The degree of polarization $P_\gamma$ is the probability ($0 \leqslant P_\gamma \leqslant 1$) of finding the state $\ket{\Phi}$ in the statistical ensemble. The density matrix is thus:
\begin{align}
    \rho_\gamma(\Phi) & = \frac{1-P_\gamma}{2} \bigg( \ket{+}\bra{+} + \ket{-}\bra{-} \bigg) + P_\gamma \ket{\Phi}\bra{\Phi} \nonumber \\
    & = \frac{1}{2} \left( I + \bm{P}_\gamma (\Phi) \cdot \bm{\sigma} \right)~,
\end{align}
where the vector $\bm{P}_\gamma(\Phi)$ depends on $P_\gamma$ and $\Phi$, $\bm{P}_\gamma = - P_\gamma \left( \cos 2\Phi, \sin 2\Phi,0 \right)$. The intensity becomes:
\begin{align}
I(\Omega,\Phi) & = I^0(\Omega) - P_\gamma I^1(\Omega) \cos 2 \Phi - P_\gamma I^2(\Omega) \sin 2 \Phi,
\end{align}
or equivalently, in the notation of Ref.~\cite{Roberts:2004mn}:
\begin{align}
I(\Omega,\Phi)  =  I^0(\Omega)\left\{1 + P_\gamma \left[ I^c(\Omega) \cos 2 \Phi + I^s(\Omega) \sin 2 \Phi\right] \right\},
\end{align}
with the obvious identification $I^{c,s} =- I^{1,2}/I^0$.

With a linearly polarized beam, the accessible moments $H^{0,1,2}$ are thus extracted from
\begin{align} 
\nonumber
H^0(LM)  &= \frac{P_\gamma}{2} \int_\circ I(\Omega,\Phi) \, 
d^L_{M0}(\theta) \cos M\phi,\\
\nonumber
H^1(LM)  &= \int_\circ  I(\Omega,\Phi) \, d^L_{M0}(\theta) \cos M\phi\, \cos 2 \Phi ,\\
\im H^2(LM) & =- \int_\circ I(\Omega,\Phi)\, d^L_{M0}(\theta) \sin M\phi \, \sin 2 \Phi,
 \label{eq:int2mom}
\end{align}
with $\int_\circ = (1/\pi P_\gamma)\int_0^\pi \sin\theta \diff \theta \int_0^{2\pi} \diff \phi \int_0^{2\pi} \diff \Phi$.

\section{Parity relations at high energies}\label{sec:parity}
In this section, we consider exchanges with spin-parity $J^{P}$, where the exchange is either natural, $P(-1)^J=+1$, or unnatural, $P(-1)^J=-1$. The properties of a particle are defined in its rest frame. In order to use the property of the exchange particle, we will use the $t$-channel frame, the rest frame of the reaction $p \bar p \to \gamma [\ell]$, where $[\ell]$ is the $\eta\pi^0$ resonance with spin $\ell$. The $t$-channel partial wave expansion reads
\begin{align}
T^{\ell,t}_{\mu_\gamma \mu_\ell; \mu_1 \mu_2} & = \sum_{J} (2J+1) a^{tJ}_{\mu_\gamma \mu_\ell; \mu_1 \mu_2}(t) d^J_{\mu \mu'}(\theta_t),
\end{align}
with $\mu = \mu_\gamma-\mu_\ell$, $\mu' = \mu_1-\mu_2$ and $\theta_t$, the scattering angle in the $t$-channel. The $t$-channel partial waves are $a^{tJ}_{\mu_\gamma \mu_\ell; \mu_1 \mu_2}(t) = \left\langle  JM \mu_\gamma \mu_\ell\right| T\left| JM \mu_1 \mu_2\right\rangle$. 
Parity imposes the relation
\begin{align}\label{eq:Atminus}
a^{tJ}_{-\mu_\gamma -\mu_\ell; \mu_1 \mu_2}(t) & =  P(-1)^J
a^{tJ}_{\mu_\gamma \mu_\ell; \mu_1 \mu_2}(t). 
\end{align}
At high energies, $\cos \theta_t \propto s$ becomes very large and the rotation function obeys the relation
\begin{align} \label{eq:dminus}
d^J_{-\mu \mu'}(\theta_t) \simeq (-1)^\mu d^J_{\mu \mu'}(\theta_t),
\end{align}
where the symbol $\simeq$ means that the relation is valid only for the leading term in $s$. In order to derive Eq.~\eqref{eq:dminus}, we use the following representation of the Wigner $d$-function~\cite{Varshalovich:1988ye}
\begin{align} \nonumber
d^J_{\mu\mu'}(\theta) & = \xi_{\mu\mu'} \left[
\frac{s! (s+m+m')!}{(s+m)!(s+m')!} \right]^{1/2} \\
& \times
\left(\sin \frac{\theta}{2} \right)^{m} \left(\cos \frac{\theta}{2} \right)^{m'}
P^{(m,m')}_s(\cos\theta),
\end{align}
with $m = |\mu-\mu'|$, $m' = |\mu+\mu'|$, $s = J - (m+m')/2$ and $\xi_{\mu\mu'} = (-1)^{(\mu'-\mu-|\mu-\mu'|)/2}$. For large value of $\cos\theta$, the leading term of the Jacobi polynomial $P^{(m,m')}_s(\cos\theta)$ leads to 
\begin{align} \nonumber
d^J_{\mu\mu'}(\theta) & \simeq (-1)^{|\mu-\mu'|/2}\xi_{\mu\mu'} \left[
\frac{s! (s+m+m')!}{(s+m)!(s+m')!} \right]^{1/2} \\
& \times \frac{\Gamma(2s+m+m'+1)}{s! \Gamma(s+m+m'+1)}
\left(\frac{\cos\theta}{2}\right)^J.
\label{eq:leadingD}
\end{align}
Under the change $\mu\to -\mu$, $m$ and $m'$ are interchanged and only the first two factors of Eq.~\eqref{eq:leadingD} change, yielding Eq.~\eqref{eq:dminus}. It is worth noting that the coefficient of the next to leading term of the Jacobi polynomial is not symmetry under the exchange $\mu\to -\mu$. The relation~\eqref{eq:dminus} thus holds only for the leading term. 

Combining the results of Eqs.~\eqref{eq:Atminus} and \eqref{eq:dminus} we obtain the relation
\begin{align}\label{eq:parityT}
T^{\ell,t}_{\mu_\gamma \mu_\ell; \mu_1 \mu_2} & \simeq P(-1)^J (-1)^{\mu_\gamma-\mu_\ell}T^{\ell,t}_{-\mu_\gamma -\mu_\ell; \mu_1 \mu_2}.
\end{align}
A similar relation can be derived for the amplitudes of the reaction $\gamma p \to [\ell] p$, by performing the boost from the $t$-channel to the helicity frame
\begin{align}
T^{\ell}_{\lambda m; \lambda_1 \lambda_2} & = e^{i \phi} \sum_{\mu_i} 
d^{\ell}_{\mu_\ell m}(\chi_\ell) 
d^{1/2}_{\mu_1 \lambda_1}(\chi_1)
d^{1/2}_{\mu_2 \lambda_2}(\chi_2)
T^{\ell,t}_{\lambda \mu_\ell; \mu_1 \mu_2}.
\label{eq:rott2s}
\end{align}  
The phase $e^{i \phi}$ and the crossing angles can be found elsewhere~\cite{Trueman:1964zzb, fox_thesis, Collins:1977jy} and do not need to be specified. Thanks to the property $d^s_{-\mu-\lambda}(\chi) = (-1)^{\mu-\lambda}d^s_{\mu\lambda}(\chi)$ and taking into account that for real photon $\lambda=\pm1$, we obtain~\cite{Cohen-Tannoudji:1968eoa}
\begin{align}\label{eq:parityHel}
T^{\ell}_{\lambda m; \lambda_1 \lambda_2} & \simeq -P(-1)^J (-1)^{m}T^{\ell,s}_{-\lambda -m; \lambda_1 \lambda_2},
\end{align}
for the helicity amplitude in the helicity frame at leading order in the energy for the exchange of particle with spin parity $J^P$. The transformation in Eq.~\eqref{eq:rott2s} being general, the relation~\eqref{eq:parityHel} holds also in every frame in which $xz$ is the reaction plane.

\section{The reflectivity basis}\label{sec:reflectivity}
We now introduce the reflectivity basis, in analogy with Ref.~\cite{Chung:1974fq}, by defining the amplitudes
\begin{align}  \label{def:Teps}
^{(\epsilon)}T^\ell_{m; \lambda_1\lambda_2} &=  \frac{1}{2} \left[ T^\ell_{+1 m; \lambda_1\lambda_2}  - \epsilon (-1)^m T^\ell_{-1 -m; \lambda_1\lambda_2}\right],
\end{align}
where, in terms of degrees of freedom, the photon helicity $\lambda$ has been traded for the reflectivity index $\epsilon = \pm$. 
The inverse relations are simply
\begin{align}\nonumber
T^\ell_{-1m; \lambda_1\lambda_2} &= (-1)^m \left[ ^{(-)}T^\ell_{-m; \lambda_1\lambda_2} -\, ^{(+)}T^\ell_{-m; \lambda_1\lambda_2} \right],\\
T^\ell_{+1m; \lambda_1\lambda_2} &= \ ^{(-)}T^\ell_{m; \lambda_1\lambda_2} +\, ^{(+)}T^\ell_{m; \lambda_1\lambda_2}.
 \label{eq:Tlam}
\end{align}
The relation~\eqref{eq:parityHel} implies that, at high energies, natural (unnatural) exchanges contributes only to the $\epsilon = +$ ($\epsilon = -$) components in the reflectivity basis. The relation between the reflectivity basis and the naturality of the exchange at high energy is the main motivation to introduce the combinations~\eqref{def:Teps}.

Parity invariance implies
\begin{align} \label{eq:parity}
^{(\epsilon)}T^\ell_{m; -\lambda_1-\lambda_2} & = \epsilon (-1)^{\lambda_1-\lambda_2}\  ^{(\epsilon)}T^\ell_{m; \lambda_1\lambda_2}.
\end{align}
We take advantage of this constraint to define 
\begin{align} \label{def:Leps}
[\ell]^{(\epsilon)}_{m;0} &=\,  ^{(\epsilon)}T^\ell_{m; ++}, & 
[\ell]^{(\epsilon)}_{m;1} &=\, ^{(\epsilon)}T^\ell_{m; +-},
\end{align}
with $[\ell] = S, P, D,\ldots$ for $\ell = 0,1,2,$ etc.
In this new basis, for each $\ell$, there are $2\times2\times(2\ell+1)$ complex partial waves $[\ell]^{(\epsilon)}_{m;k}$ with $\epsilon = \pm$, $k=0,1$ and $m = -\ell, \ldots, \ell$. It is worth noticing that, in the reflectivity basis for photoproduction, $m$ takes positive and negative values. {\it A contrario}, in the reflectivity basis for spinless beam $m$ is only positive~\cite{Chung:1974fq}. 

Another advantage of this basis is to diagonalize the spin density matrix element in the $\epsilon$ space.
In order to obtain this result, we first perform the summation over the photon helicities $\lambda = \pm 1$ in the definitions of the spin density matrices, Eqs.~\eqref{eq:rho}. Then we substitute the amplitudes with photon helicities by the reflectivity basis using the definitions in Eqs.~\eqref{eq:Tlam}. We finally use to the parity relation in Eq.~\eqref{eq:parity} to recast the interference terms as
\begin{align} \label{eq:sumHel12}
\sum_{\lambda_1 \lambda_2} \, ^{(\epsilon)}T^{\ell}_{m; \lambda_1\lambda_2} 
\, ^{(\epsilon')}T^{\ell' *}_{m'; \lambda_1\lambda_2} & = 2 \delta_{\epsilon,\epsilon'}\sum_{k} \,  [\ell]^{(\epsilon)}_{m;k} [\ell']^{(\epsilon)*}_{m';k}.
\end{align}
The interference between different $\epsilon$ thus vanishes and the intensities, moments and SDME are split into an incoherent sum over the different reflectivity components. For the moments we write
\begin{align} \label{eq:Hpm}
H^{\alpha}(LM) &=  \, ^{(+)}H^{\alpha}(LM) + \, ^{(-)}H^{\alpha}(LM)~,
\end{align}
and similarly for the density matrices
\begin{align}
\rho^{\alpha, \ell \ell'}_{mm'} &=  \, ^{(+)}\rho^{\alpha, \ell \ell'}_{mm'} + \, ^{(-)}\rho^{\alpha, \ell \ell'}_{mm'}~.
\end{align}
With this convention, the explicit expressions for the spin density matrices in terms of partial waves read
\bsub \label{eq:rho-pw}
\begin{align} \nonumber
\, ^{(\epsilon)}\rho^{0, \ell \ell'}_{mm'} & = \kappa \sum_{k} \Big( [\ell]^{(\epsilon)}_{m;k} [\ell']^{(\epsilon)*}_{m';k}  \\
&\quad \qquad + (-1)^{m-m'} [\ell]^{(\epsilon)}_{-m;k}[\ell']^{(\epsilon)*}_{-m';k} \Big)~, \\  \nonumber
\, ^{(\epsilon)}\rho^{1, \ell \ell'}_{mm'} & = -\epsilon \kappa\sum_{k} \Big( (-1)^m  [\ell]^{(\epsilon)}_{-m;k} [\ell']^{(\epsilon)*}_{m';k} \\
& \qquad \quad +(-1)^{m'}  [\ell]^{(\epsilon)}_{m;k} [\ell']^{(\epsilon)*}_{-m';k} \Big)~, 
\\ \nonumber
\, ^{(\epsilon)}\rho^{2, \ell \ell'}_{mm'} & = -i\epsilon \kappa \sum_k \Big( (-1)^m [\ell]^{(\epsilon)}_{-m;k} [\ell']^{(\epsilon)*}_{m';k} \\
& \qquad \quad -(-1)^{m'}  [\ell]^{(\epsilon)}_{m;k} [\ell']^{(\epsilon)*}_{-m';k}  \Big)~, \\ \nonumber
\, ^{(\epsilon)}\rho^{3, \ell \ell'}_{mm'} & =  \kappa \sum_k \Big( [\ell]^{(\epsilon)}_{m;k} [\ell']^{(\epsilon)*}_{m';k} \\
&\qquad \quad - (-1)^{m-m'} [\ell]^{(\epsilon)}_{-m;k} [\ell']^{(\epsilon)*}_{-m';k} \Big)~.
\end{align} \esub
Equations~\eqref{eq:rho-pw} are useful to express moments $H^{\alpha}(LM)$ in terms of partial waves. From Eqs.~\eqref{eq:rho-pw} we can also extract the relations
\bsub \label{eq:rho_good_nat}
\begin{align} 
\, ^{(\epsilon)}\rho^{1, \ell \ell'}_{mm'} & = -\epsilon (-1)^{m} \, ^{(\epsilon)}\rho^{0, \ell \ell'}_{-mm'}, \\
\, ^{(\epsilon)}\rho^{3, \ell \ell'}_{mm'} & =i \epsilon (-1)^{m} \, ^{(\epsilon)}\rho^{2, \ell \ell'}_{-mm'}.
\end{align} \esub
From the knowledge of the spin density matrix elements $\rho^{\alpha, \ell \ell'}_{mm'}$ one can reconstruct the good reflectivity elements {\it via}
\bsub\begin{align}
\, ^{(\epsilon)}\rho^{0,\ell \ell'}_{mm'}  & =  \frac{1}{2}\left(\rho^{0,\ell \ell'}_{mm'} - \epsilon (-1)^m \rho^{1, \ell \ell'}_{-m m'} \right),  \\
\, ^{(\epsilon)}\rho^{3,\ell \ell'}_{mm'}  & =  \frac{1}{2}\left(\rho^{3,\ell \ell'}_{mm'} +i \epsilon (-1)^m \rho^{2, \ell \ell'}_{-m m'} \right).
\end{align}\esub
In the case of the dominance of a single partial wave, SDME can be extracted from the angular distribution of the data and the formalism presented is equivalent to the one introduced  in Ref.~\cite{Schilling:1969um}. When more than one wave contribute to the partial wave expansion, SDME cannot be isolated, and only moments can be extracted.

The intensities are also an incoherent sum over the reflectivities. In order to express the intensities int term of the partial waves in the reflectivity basis, we introduce the quantities
\bsub \begin{align}
    U^{(\epsilon)}_{k}(\Omega) & = \sum_{\ell, m} [\ell]_{m;k}^{(\epsilon)} Y_\ell^{m}(\Omega)~,\\
    \widetilde U^{(\epsilon)}_{k}(\Omega) & = \sum_{\ell, m} [\ell]_{m;k}^{(\epsilon)} \left[Y_\ell^{m}(\Omega)\right]^*~.
\end{align}\esub
The quantities $U^{(\epsilon)}_{k}(\Omega)$ and $\widetilde U^{(\epsilon)}_{k}(\Omega)$ are not helicity amplitudes. They arise when the parity relations are used to replace the sum over nucleon helicities by the sum over $k$, as in Eq.~\eqref{eq:sumHel12}. The intensities can be expressed by 
\bsub
\begin{align}
    I^0(\Omega) & =\phantom{-}\kappa \sum_{\epsilon,k} |U^{(\epsilon)}_{k}(\Omega)|^2 + |\widetilde U^{(\epsilon)}_{k}(\Omega)|^2~, \\
    I^1(\Omega) & = -\kappa \sum_{\epsilon,k} 2 \epsilon \re \left(U^{(\epsilon)}_{k}(\Omega) \left[\widetilde U^{(\epsilon)}_{k}(\Omega)\right]^* \right) ~, \\
    I^2(\Omega) & = -\kappa \sum_{\epsilon,k} 2 \epsilon \im \left(U^{(\epsilon)}_{k}(\Omega) \left[\widetilde  U^{(\epsilon)}_{k}(\Omega)\right]^* \right)~,\\
    I^3(\Omega) & = \phantom{-}\kappa \sum_{\epsilon,k} |U^{(\epsilon)}_{k}(\Omega)|^2 - |\widetilde U^{(\epsilon)}_{k}(\Omega)|^2~.
\end{align} \esub
For a linearly beam, one can write the full intensity as
\begin{align} \nonumber
I(\Omega,\Phi)  = 2 \kappa & \sum_k 
(1+P_\gamma) \left| [\ell]_{m;k}^{(+)} \re Z_{\ell}^m(\Omega,\Phi) \right|^2 \\ \nonumber
& + (1-P_\gamma) \left| [\ell]_{m;k}^{(+)} \im Z_{\ell}^m(\Omega,\Phi) \right|^2 \\ \nonumber
& + (1-P_\gamma) \left| [\ell]_{m;k}^{(-)} \re Z_{\ell}^m(\Omega,\Phi) \right|^2 \\
&+ (1+P_\gamma) \left| [\ell]_{m;k}^{(-)} \im Z_{\ell}^m(\Omega,\Phi) \right|^2.
\label{eq:full_int}
\end{align}
In Eq.~\eqref{eq:full_int}, we have defined the quantity $Z_{\ell}^m(\Omega,\Phi) = Y_\ell^m(\Omega) e^{-i\Phi}$, such that
\bsub\begin{align}
\re Z_\ell^m(\Omega,\Phi) & = \sqrt{\frac{2\ell+1}{4\pi}} d^\ell_{m0}(\theta) \cos (m\phi-\Phi), \\
\im Z_\ell^m(\Omega,\Phi) & = \sqrt{\frac{2\ell+1}{4\pi}} d^\ell_{m0}(\theta) \sin (m\phi-\Phi)
\end{align}\esub
Finally let us prove \eqref{eq:H2H1}. We use Eqs.~\eqref{eq:rho-pw} to express the difference $ \Delta \equiv  \im H^2(LM) + H^1(LM)$, as
\begin{align}\nonumber
\Delta
& = 2\kappa\sum_{k, \epsilon} \sum_{\substack{\ell \ell' \\ m m'} } \left(  \frac{2\ell'+1}{2\ell+1} \right)^{1/2} 
C^{\ell 0}_{\ell' 0 L0} C^{\ell m}_{\ell'm' LM} \\
& \qquad \qquad \qquad \qquad \times   \epsilon (-1)^m [\ell]^{(\epsilon)}_{-m;k} [\ell']^{(\epsilon)*}_{m';k}.
\label{eq:proof}
\end{align}
Since the basis includes only positive spin projection components, $\Delta$ vanishes unless the summation indices satisfy $m \leqslant 0$ and $m' = m-M \geqslant 0$. These conditions are incompatible with $M \geqslant 1$. Consequently we obtain the condition
\begin{align}\label{eq:H2H12}
\im H^2(LM) & = -H^1(LM), & \text{for } M \geqslant 1.
\end{align}
for any wave set restricted to only positive $m$, and thus for our wave set~\eqref{eq:waves}. 

From an experimental perspective, the moments are extracted from the angular distribution, \textit{cf.} Eqs.~\eqref{eq:int2mom}, without assuming a particular wave content. If the experimentally extracted moments were not to satisfy the condition in Eq.~\eqref{eq:H2H1}, it would indicate that negative $m$ components (in the reflectivity basis) are required for a proper description of the two meson system.

\section{\boldmath Moments with $S$, $P$ and $D$ waves} \label{sec:app}
We restrict the wave set to only $S$-, $P$- and $D$-waves with only positive $m$ components. 
The moments $H^3(LM)$ are not accessible with a linearly polarized beam and we have already proven that $\im H^2(LM)= -H^1(LM)$, \textit{cf}. Eq.~\eqref{eq:proof}. Our basis~\eqref{eq:waves} include only positive reflectivity components,  the relevant moments are thus $H^{0,1}(LM) =\ ^{(+)}H^{0,1}(LM)$. 
We do not include the phase space factor $\kappa$ to simplify the equations. 
In terms of partial waves, the moments for $L = 0,1,2$ are:
\begingroup
\allowdisplaybreaks
\bsub \label{eq:appmom1}
\begin{align}
H^{0}(00) &= H^{1}(00) + 2 \left[ |P_1^{(+)}|^2  + |D_1^{(+)}|^2  + |D_2^{(+)}|^2   \right]~, \\
H^{1}(00) &= 2 \left[ |S_0^{(+)}|^2  + |P_0^{(+)}|^2 + |D_0^{(+)}|^2  \right]~, \\
H^{0}(10) &= H^{1}(10) + \frac{4}{\sqrt{5}}  \re (P_1^{(+)} D_1^{(+)*})~,  \\
H^{1}(10) &= \frac{8}{\sqrt{15}} \re (P_0^{(+)} D_0^{(+)*}) + \frac{4}{\sqrt{3}}  \re (S_0^{(+)} P_0^{(+)*}) ~, \\
H^{0}(11) &= H^{1}(11) + 2\sqrt{\frac{2}{5}} \re (P_1^{(+)} D_2^{(+)*})~, \\ \nonumber
H^{1}(11) &= \frac{2}{\sqrt{5}} \re (P_0^{(+)} D_1^{(+)*}) - \frac{2}{\sqrt{15}}\re (P_1^{(+)} D_0^{(+)*})  \\
&+\frac{2}{\sqrt{3}}\re (S_0^{(+)} P_1^{(+)*}) ~, \\ 
H^{0}(20) &= H^{1}(20) - \frac{2}{5} |P_1^{(+)}|^2  + \frac{2}{7}  |D_1^{(+)}|^2 -\frac{4}{7}  |D_2^{(+)}|^2~, \\
H^{1}(20) &=  \frac{4}{5} |P_0^{(+)}|^2 + \frac{4}{7}  |D_0^{(+)}|^2 + \frac{4}{\sqrt{5}}  \re (S_0^{(+)} D_0^{(+)*})~, \\ 
H^{0}(21) &= H^{1}(21) + \frac{2}{7}\sqrt{6} \re (D_1^{(+)} D_2^{(+)*})~,  \\ 
\nonumber
H^{1}(21) &= \frac{2}{\sqrt{5}}  \re (S_0^{(+)} D_1^{(+)*})  +\frac{2 \sqrt{3}}{5} \re (P_0^{(+)} P_1^{(+)*}) \\
&  + \frac{2}{7}  \re (D_0^{(+)} D_1^{(+)*})~, \\ 
H^{0}(22) &=\frac{2}{\sqrt{5}} \re (S_0^{(+)} D_2^{(+)*}) - \frac{4}{7}  \re (D_0^{(+)} D_2^{(+)*})~, \\ 
H^{1}(22) &= H^{0}(22) +\frac{\sqrt{6}}{7} |D_1^{(+)}|^2 + \frac{\sqrt{6}}{5} |P_1^{(+)}|^2,
\end{align}\esub
\endgroup
and for $L = 3,4$
\bsub  \label{eq:appmom2}\begin{align}
H^{0}(30) &= H^{1}(30) - \frac{12}{7\sqrt{5}}  \re( P_1^{(+)} D_1^{(+)*})~, \\
H^{1}(30) &= \frac{12}{7} \sqrt{\frac{3}{5}} \re( P_0^{(+)} D_0^{(+)*})~, \\
H^{0}(31) &= H^{1}(31) - \frac{2}{7} \sqrt{\frac{3}{5}} \re( P_1^{(+)} D_2^{(+)*})~, \\
H^{1}(31) &= \frac{4}{7} \sqrt{\frac{6}{5}} \re( P_0^{(+)} D_1^{(+)*}) + \frac{6}{7} \frac{\sqrt{2}}{5} \re( P_1^{(+)} D_0^{(+)*})~, \\
H^{0}(32) &= H^{1}(32)  - \frac{2}{7} \sqrt{6} \left[  \re( P_1^{(+)} D_1^{(+)*}) \right]~, \\
H^{1}(32) &= \frac{2}{7} \sqrt{3} \left[  \re( P_0^{(+)} D_2^{(+)*})+ \sqrt{2}  \re( P_1^{(+)} D_1^{(+)*}) \right]~, \\
H^{0}(33) &= 0~, \\
H^{1}(33) &= \frac{6}{7}   \re( P_1^{(+)} D_2^{(+)*})~,  \\
H^{0}(40) &= H^{1}(40) - \frac{2}{21} \left[ 4 |D_1^{(+)} |^2 - |D_2^{(+)} |^2\right]~, \\
H^{1}(40) &= \frac{4}{7} |D_0^{(+)} |^2~, \\
H^{0}(41) &= H^{1}(41)  - \frac{2}{21} \sqrt{5}  \re(D_1^{(+)} D_2^{(+)*} )~, \\
H^{1}(41) &= \frac{2}{7} \sqrt{\frac{10}{3}} \re(D_0^{(+)} D_1^{(+)*} )~,  \\
H^{0}(42) &= -\frac{2}{7} \sqrt{\frac{5}{3}} \re(D_0^{(+)} D_2^{(+)*} )~,  \\
H^{1}(42) &= H^{0}(42)  + \frac{2\sqrt{10}}{21} |D_1^{(+)}|^2~, \\
H^{0}(43) &=  H^{(0)}(44) = 0~,\\
H^{1}(43) &= \frac{2}{3}\sqrt{\frac{5}{7}}  \re(D_1^{(+)} D_2^{(+)*} )~,   \\
H^{1}(44) &= \frac{1}{3}\sqrt{\frac{10}{7}}  |D_2^{(+)}|^2~. 
\end{align} \esub


\end{document}